\documentclass[pdflatex,sn-nature]{sn-jnl}

\usepackage{graphicx}%
\usepackage{amsmath,amssymb,amsfonts}%
\usepackage[title]{appendix}%
\usepackage[table]{xcolor}%
\usepackage{booktabs}%
\usepackage{multirow}%

\newcommand{\metricdown}[1]{\textbf{#1} $\downarrow$}
\newcommand{\metricup}[1]{\textbf{#1} $\uparrow$}

\begin{document}

\title[FARM: Foundational Aerial Radio Map for Intelligent Low-Altitude Networking]{FARM: Foundational Aerial Radio Map for Intelligent Low-Altitude Networking}

\author[1]{\fnm{Shijian} \sur{Gao}}
\author[1]{\fnm{Jiahui} \sur{Liang}}
\author[1]{\fnm{Yifeng} \sur{Yuan}}
\author[1]{\fnm{Wenlihan} \sur{Lu}}
\author[1]{\fnm{Guobin} \sur{Shen}}
\author[1]{\fnm{Liuqing} \sur{Yang}}

\affil[1]{\orgname{The Hong Kong University of Science and Technology (Guangzhou)}, \orgaddress{\city{Guangzhou}, \postcode{511400}, \country{P.R. China}}}



\abstract{
Precise aerial radio environment characterization is vital for low-altitude airspace planning. However, existing datasets and construction methods lack the high-resolution granularity required for complex aerial spaces, particularly failing to capture spatial variations across both horizontal and vertical dimensions. To address these gaps, this paper introduces FARM, a pioneering foundation model for unified aerial radio map (ARM) construction. FARM is supported by our newly curated, high-granularity full-domain ARM dataset, which features multi-band and multi-antenna configurations, effectively filling a critical void in comprehensive low-altitude radio data. Structurally, FARM leverages a masked autoencoder to extract deep latent representations of the aerial radio environment, which subsequently guide a diffusion-based decoder to synthesize high-fidelity signal distributions through only a few iterative refinement steps. Benefiting from this design, the architecture seamlessly accommodates both condition-based and condition-free ARM construction, providing robust support for diverse signal and environmental priors. Extensive experiments demonstrate that FARM significantly outperforms state-of-the-art benchmarks while exhibiting strong cross-scenario generalization. Crucially, we validate the transferability of FARM on a real-world dataset collected from field tests, proving its robust deployment capability. Ultimately, FARM serves as a foundational infrastructure for the low-altitude economy by enabling autonomous aerial logistics and intelligent urban networking.
}


\maketitle

\section{Introduction}

The rapid evolution of low-altitude networks necessitates precise aerial radio environment modeling to ensure reliable communication, effective network planning, and intelligent service deployment \cite{JYJW_npjWT_2026, GYHL_IoTM_2026}. To support this demand, the aerial radio map (ARM) serves as a fundamental tool to characterize the spatial distribution of pathloss from transmitters to any location in aerial space \cite{SSDS_ComST_2025}. However, obtaining high-quality ARMs at scale remains a significant challenge \cite{ZCGC_ComST_2024}, because physical measurements are prohibitively expensive for large-scale deployment, whereas high-fidelity ray-tracing simulations incur unsustainable computational and time overhead \cite{HAGW_ComST_2019}.

As an efficient alternative, many works have focused on constructing the entire radio map from partial received signal strength (RSS) observations or radio environmental priors \cite{HASJ_scirep_2025}. The priors may include environmental structure, such as terrain/building layouts, as well as radio configurations, such as base station (BS) location and transmission parameters. According to whether these priors is used, radio map construction approaches can be categorized into two paradigms: condition-free methods, which construct radio maps using only RSS samples, and condition-based methods, which construct radio maps guided by priors. As a condition-free method, interpolation techniques have been frequently used over the decades \cite{SaFu_TCCN_2017, SuCh_TSP_2022}. With the development of artificial intelligence (AI), deep learning approaches such as autoencoders (AEs) and U-Nets have gained increasing employment \cite{KHC_ICC_2021, CQG_WCL_2025}. Nevertheless, the condition-free methods typically require high sampling rates to achieve reasonable accuracy. This creates a bottleneck in ``few-sample'' environments, particularly in regions that include low-altitude no-fly zones \cite{TZQS_scirep_2023}. In this regard, condition-based methods offer a promising way to alleviate this issue by leveraging environmental structure in ultra-sparse observation scenarios \cite{LZLZ_JSAC_2025, LLZZ_JSAC_2026}. Along this route, recent literature has largely framed radio map construction as an image-to-image translation task and has additionally leveraged radio configurations to enhance the precision \cite{LYKC_TWC_2021, JCD_TWC_2026}. Advanced generative frameworks have recently been adopted to construct radio maps \cite{ZWD_IoTJ_2023}. Notably, \cite{WTCY_TCCN_2025} first uses a diffusion model to generate radio maps merely via radio environmental priors. Therefore, condition-based methods have two input cases: condition-and-sample input and condition-only input. For three-dimensional (3D) aerial scenarios, current works often adopt a layer-by-layer construction approach from sliced building maps \cite{HHCW_TVT_2023, ZFWL_WCL_2025}. As revealed by these existing works, condition-based methods rely heavily on high-resolution environmental maps and precise BS configurations. This is a critical limitation since such information is often unavailable in practice due to security restrictions or privacy concerns, where condition-free methods are expected to serve as a fallback solution. However, existing methods are typically tailored to fixed paradigms, making it difficult to seamlessly support unified ARM construction under varying priors. Consequently, separate models have to be trained for different inputs, which increases training and deployment overhead. Moreover, both condition-free and condition-based techniques often struggle with cross-scenario generalization, requiring additional adaptation for each new environment. This limitation becomes particularly impractical for dynamic low-altitude networks with heterogeneous BS configurations.

In the meantime, recent advances in foundation models have redefined AI \cite{abramson2024alphafold3, moor2023gmai}, opening new possibilities to address the problems outlined above. Trained with self-supervision on massive datasets, foundation models \cite{he2025generalized, pai2024foundation} learn transferable representations that can be adapted to a wide range of downstream tasks through fine-tuning or zero-shot inference. This general-purpose learning paradigm has repeatedly outperformed conventional task-specific designs. Within the wireless research field, early studies \cite{LGLC_scichina_2025, ZGC_WCL_2026, YWZL_ICC_2026, wifocp2026twc, LLGC_arXiv_2026} have begun to demonstrate the potential of foundation models. Up to date, however, no existing architecture is capable of addressing unified ARM construction with high-resolution granularity. In addition, foundation models hinge upon high-resolution, large-scale ARM datasets with diverse transmission configurations. As of today, the latter are essentially nonexistent, rendering the foundation model development impossible. This is corroborated by the list of pertinent datasets in Table~\ref{tab:dataset_comparison}. More specifically, most public datasets are designed primarily for terrestrial scenarios, although datasets such as RadioGAT \cite{LZLL_TWC_2024} and RMDirectionalBerlin \cite{JCD_TWC_2026} offer various frequency and antenna configurations. Even datasets with multiple height levels, such as UrbanRadio3D \cite{WZQC_TNSE_2026} and SpectrumNet \cite{ZJLF_TCCN_2025}, lack sufficient coverage of the aerial environment and feature small map grid sizes that preclude fine-grained modeling of signal propagation over expansive regions.

To fill these gaps, this paper introduces ARM-Omni, the first large-scale low-altitude domain ARM dataset featuring high-resolution, multi-band and multi-antenna patterns. Unlike existing datasets that treat distinct heights as isolated two-dimensional (2D) slices, ARM-Omni captures the continuous vertical evolution of radio propagation across 3D uniformly spaced altitude levels, allowing foundation models to exploit cross-layer statistical dependencies. With the support of ARM-Omni, FARM is engineered to facilitate two complementary capabilities: i) generalizable aerial radio representation understanding from sparse observations; and ii) high-fidelity ARM generation from radio environmental priors. The FARM architecture combines a masked autoencoder (MAE) for deep representation learning \cite{HCXL_CVPR_2022} with a diffusion-based decoder for high-fidelity generative refinement \cite{HJA_Neuripe_2020}, where voxel-space alignment bridges the two components. A specialized two-stage training strategy, consisting of self-supervised pretraining followed by generative fine-tuning, enables FARM to adapt the use of its encoder and decoder for different ARM construction paradigms.

\begin{table}[t]
\centering
\caption{Comparison of ARM datasets (``--'' and ``$\blacktriangle$'' denote not disclosed and limited support, respectively. The \colorbox{red!15}{light red} shading indicates the best results).}
\label{tab:dataset_comparison}
\tiny
{\setlength{\tabcolsep}{0pt}
\renewcommand{\arraystretch}{1.5}
\begin{tabular}{>{\raggedright\arraybackslash}p{0.18\columnwidth}
                >{\centering\arraybackslash}p{0.115\columnwidth}
                >{\centering\arraybackslash}p{0.145\columnwidth}
                >{\centering\arraybackslash}p{0.13\columnwidth}
                >{\centering\arraybackslash}p{0.24\columnwidth}
                >{\centering\arraybackslash}p{0.10\columnwidth}
                >{\centering\arraybackslash}p{0.145\columnwidth}@{}}
\toprule
\textbf{Properties} & \textbf{ARM-Omni (Ours)} & \textbf{UrbanRadio3D} & \textbf{SpectrumNet} & \textbf{RMDirectionalBerlin} & \textbf{RadioGAT} & \textbf{RadioMapSeer} \\
\midrule
Dataset Size & \cellcolor{red!15}\textbf{26.4M} & 11.2M & 300K & 75K & 21K & 56K \\
Number of BSs & \cellcolor{red!15}\textbf{$\geq 10^2$} & $\geq 10^2$ & $\leq 10$ & $\leq 10$ & $\leq 10$ & $\geq 10$ \\
Resolution & \cellcolor{red!15}\textbf{1000$\times$1000} & 256$\times$256 & 128$\times$128 & 256$\times$256 & 200$\times$200 & 256$\times$256 \\
Coverage Area (m$^2$) & \textbf{1000$\times$1000} & 256$\times$256 & \cellcolor{red!15}\textbf{1280$\times$1280} & -- & 1000$\times$1000 & 256$\times$256 \\
Rx Height Range (m) & \cellcolor{red!15}\textbf{5--150} & 1--20 & 1.5, 30, 200 & 1.5 & -- & 1.5 \\
Rx Height Levels & \cellcolor{red!15}\textbf{30} & 10 & 3 & 1 & 1 & 1 \\
Frequencies & \cellcolor{red!15}\textbf{7} & 1 & 5 & 1 & 5 & 1 \\
Antenna Patterns & \cellcolor{red!15}\textbf{4} & 1 & 1 & 3 & 1 & 1 \\
Aerial Measurements & \cellcolor{red!15}\textbf{$\checkmark$} & $\blacktriangle$ & $\blacktriangle$ & $\times$ & $\times$ & $\times$ \\
\bottomrule
\end{tabular}}
\end{table}

Our specific contributions can be summarized as follows:

\begin{itemize}
    \item We propose FARM, a pioneering foundation model that synergistically couples an MAE with a diffusion mechanism. By design, FARM seamlessly accommodates both condition-based and condition-free ARM construction, providing unprecedented flexibility across diverse signal and environmental priors with strong cross-scenario generalization.
    
    \item We introduce ARM-Omni, the first large-scale, high-resolution dataset featuring multi-band and multi-antenna configurations specifically tailored for low-altitude spaces. ARM-Omni captures fine-grained spatial variations across both horizontal and vertical dimensions, effectively filling a critical void in comprehensive 3D air-ground propagation modeling.
    
    \item We conduct extensive experiments demonstrating that FARM significantly outperforms state-of-the-art (SOTA) benchmarks in construction accuracy while maintaining superior generalization across unseen, heterogeneous BS configurations. Crucially, we validate the transferability of FARM on a real-world dataset collected from field tests, demonstrating its robust deployment capability in practical low-altitude applications.
\end{itemize}

\noindent\textit{Notation:} $a$, $\mathbf{a}$, $\mathbf{A}$, $\mathcal{A}$ and $\varnothing$ represent a scalar, a vector, a matrix, a set, and an empty set, respectively. For a matrix $\mathbf{A}$, $\mathbf{A}[i,j]$ and $\mathbf{A}[i,:]$ denote the entry in the $i$-th row and the $j$-th column and the $i$-th row, respectively. $(\cdot)^{\mathrm{T}}$, $\|\cdot\|_2$, $\mathbb{E}(\cdot)$ and $\lfloor\cdot\rfloor$ represent the transpose, 2-norm, expectation, and floor operation, respectively. $\odot$ and $\oplus$ denote the Hadamard product and the concatenation operation. $\mathcal{N}(\boldsymbol{\mu},\mathbf{\Sigma})$ denotes the Gaussian distribution with mean $\boldsymbol{\mu}$ and covariance $\mathbf{\Sigma}$, and $\mathbf{I}$ represents the identity matrix.

\section{System Model and Problem Formulation}
\label{secII}

This paper considers an aerial radio environment discretized into a three-dimensional (3D) voxel grid of size $L \times W \times H$. The set of voxel indices is denoted by $\mathcal{V}$, where $\mathbf{x}_i \in \mathbb{R}^3$ represents the center coordinates of voxel $i \in \mathcal{V}$. A single-antenna BS is deployed at coordinates $\mathbf{p}_{tx} = (l_{tx}, w_{tx}, h_{tx})$ with a transmit power $P_{tx}$ to serve the specified region. The transmission parameters of BS are defined as $\mathbf{c} = (f_c, \mathbf{a})$, where $f_c$ denotes the carrier frequency and $\mathbf{a} = (\tau, \phi_{tx}, \theta_{tx})$ represents the antenna settings. Specifically, $\tau \in \{\text{iso}, \text{dir}\}$ specifies the antenna type, while $\phi_{tx}$ and $\theta_{tx}$ denote the boresight azimuth and elevation angles, respectively.

\begin{figure*}[!t]
\centering
\includegraphics[width=\textwidth]{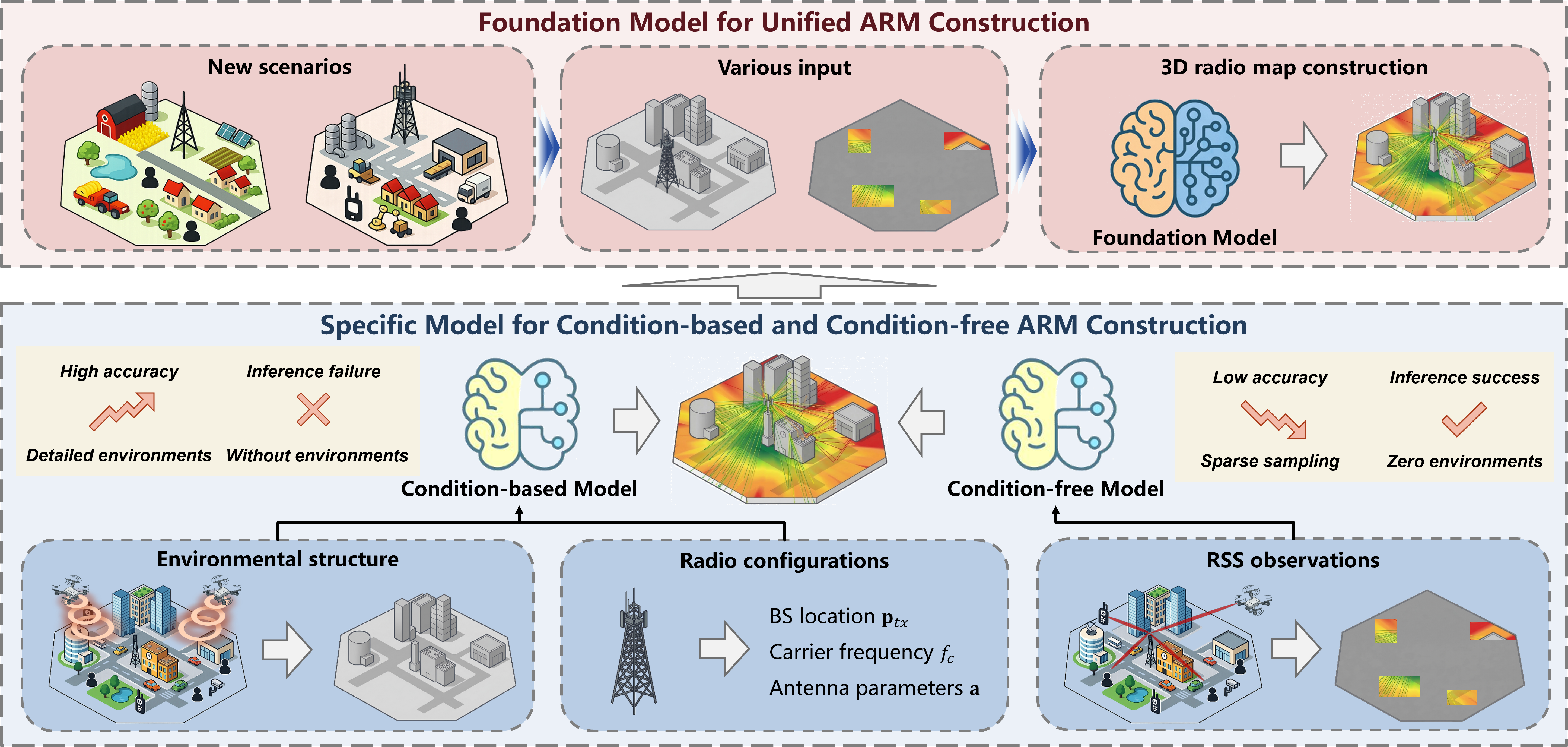}
\caption{Foundation model versus specific model for ARM construction.}
\label{fig:system_model}
\end{figure*}

The transmit antenna gain for the $i$-th voxel is defined as $g_i = \pi(\phi_i, \theta_i, \tau)$, where $\pi(\cdot)$ denotes the radiation pattern parameterized by the antenna type and orientation. Let $\ell_i$ represent the effective large-scale propagation attenuation from the BS to voxel $i$. The corresponding large-scale RSS is expressed as:
\begin{equation}
    r_i = P_{tx} + g_i - \ell_i, \quad i \in \mathcal{V}.
\end{equation}
Accordingly, the ARM is defined as $\mathbf{R} \in \mathbb{R}^{L \times W \times H}$ by aggregating the RSS values across the entire voxel grid. The objective of this work is to develop a neural network $\mathcal{G}(\cdot)$, parameterized by $\mathbf{\Theta}$, capable of constructing the ARM $\hat{\mathbf{R}}$ in both condition-based and condition-free paradigms. For the condition-based paradigm, the ARM construction is guided by radio environmental priors, including the BS location, transmission parameters and a 3D building map $\mathbf{B} \in \mathbb{R}^{L \times W \times H}$. In the condition-free mode, such priors are assumed to be unavailable, and the ARM is constructed solely based on a set of observed RSS samples from $S$ voxels, denoted as $\mathcal{R} = \{r_s\}_{s=1}^{S}$. Accordingly, the unified ARM construction problem is formulated as:

\begin{subequations}
\begin{align}
\label{eq:prob1}
\min_{\mathbf{\Theta}} \quad
& \|\mathbf{R}-\hat{\mathbf{R}}\|_2^2\\
\text{s.t.}\quad
& \hat{\mathbf{R}} = {\mathcal{G}}(\mathbf{B},\mathbf{c}, \mathbf{p}_{tx}, \mathcal{R}).
\end{align}
\end{subequations}
As shown in Fig.~\ref{fig:system_model}, traditional methodologies generally fail to support both condition-based and condition-free construction within a single, unified framework. Furthermore, existing approaches often lack the flexibility to accommodate heterogeneous system configurations, such as varying coverage dimensions $L \times W \times H$, diverse carrier frequencies $f_c$, and complex antenna parameters $\mathbf{a}$, all of which are essential for robust performance in the low-altitude economy.

\section{Architecture and Pipelines for FARM}

To integrate representation learning and generative modeling for unified ARM construction, we propose FARM, a foundation model consisting of an MAE-based radio encoder and a diffusion-based map decoder. These two modules are aligned in voxel space and jointly optimized in velocity space, allowing for adaptive execution across different construction modes. Through this architecture, FARM achieves SOTA accuracy in both condition-free and condition-based ARM construction tasks, while demonstrating robust generalization across unseen operating coverage dimensions, frequencies, and antenna configurations.

\subsection{Alignment of MAE and Diffusion}
\label{secIII}

Despite their distinct learning objectives, we realize that MAE and diffusion models share the same fundamental reconstruction mechanism. Rather than treating them as a simple combination of two separate frameworks, FARM unifies them as an elegant, intentional duality. Standard diffusion architectures typically generate data entirely from scratch, relying solely on radio environmental priors (condition-only), whereas standard MAEs are designed strictly to interpolate missing spatial gaps from partial observations (condition-free). This unified view allows FARM to adaptively coordinate its radio encoder and map decoder according to the available inputs.

Methodologically, the two models are unified through a shared voxel-space recovery objective \cite{LiHe_arXiv_2025}. In MAE, the input is divided into patches, and a large portion is randomly masked. The encoder processes only the visible patches to produce latent representations, which are then used by the decoder to reconstruct the missing content in the voxel space. Similarly, we optimize the diffusion decoder to directly predict the clean sample $\hat{\mathbf{R}}$ during denoising rather than the noise itself. By grounding both representation learning and map generation in a common voxel-space target, the unified design remains coherent and computationally efficient. 

To implement this, the MAE $\mathcal{G}^{\mathrm{enc}}(\cdot)$ utilizes a vanilla Vision Transformer (ViT) backbone \cite{Alex_ICLR_2021}, while the diffusion decoder $\mathcal{G}^{\mathrm{dec}}(\cdot)$ is built upon a Diffusion Transformer (DiT) architecture \cite{PEXI_ICCV_2022}. To reduce inference overhead, we formulate the generation process of the decoder as an Ordinary Differential Equation (ODE) following the flow-matching paradigm \cite{LCBM_ICLR_2022}. Let $\boldsymbol{\epsilon} \sim \mathcal{N}(\mathbf{0}, \mathbf{I})$ denote a standard Gaussian noise sample. A noisy intermediate sample $\mathbf{Z}_t$ at time $t \in [0,1]$ is constructed from the original input $\mathbf{R}$ via linear interpolation:
\begin{equation}
    \label{eq5}
    \mathbf{Z}_t = t\mathbf{R} + (1-t)\boldsymbol{\epsilon}.
\end{equation}
Under this probability path, the target velocity is $\mathbf{v} = \frac{d\mathbf{Z}_t}{dt} = \mathbf{R} - \boldsymbol{\epsilon}$. Since $\mathbf{v}$ is constant for each pair $(\mathbf{R}, \boldsymbol{\epsilon})$, it is significantly easier to learn than stochastic formulations. To satisfy the voxel-space prediction requirement $\hat{\mathbf{R}} = \mathcal{G}^{\mathrm{dec}}(\mathbf{Z}_t, t)$, we transform the output of the diffusion decoder into the velocity space. Specifically, by rearranging Eq. \eqref{eq5}, the predicted noise is recovered as $\hat{\boldsymbol{\epsilon}} = \frac{\mathbf{Z}_t - t\hat{\mathbf{R}}}{1-t}$. The corresponding predicted velocity is then computed as:
\begin{equation}
    \label{eq9}
    \hat{\mathbf{v}} = \hat{\mathbf{R}} - \hat{\boldsymbol{\epsilon}} = \frac{\mathcal{G}^{\mathrm{dec}}(\mathbf{Z}_t, t) - \mathbf{Z}_t}{1-t}.
\end{equation}
This yields the ODE $\frac{d\mathbf{Z}_t}{dt} = \hat{\mathbf{v}}$, which is solved numerically during inference. Accordingly, the model is trained with the flow-matching objective:
\begin{equation}
\label{eq6}
\mathcal{L} = \mathbb{E}_{t,\mathbf{R},\boldsymbol{\epsilon}} \left\| \hat{\mathbf{v}} - \mathbf{v} \right\|_2^2 = \mathbb{E}_{t,\mathbf{R},\boldsymbol{\epsilon}} \left\| \frac{\hat{\mathbf{R}} - \mathbf{Z}_t}{1-t} - (\mathbf{R} - \boldsymbol{\epsilon}) \right\|_2^2.\end{equation}

This formulation bridges voxel-space supervision with ODE-based generation, allowing FARM to predict high-fidelity ARMs while maintaining the mathematical rigor of flow-matching optimization. The resulting model functions as a deep interpolation engine when utilizing the ViT-based radio encoder to interpret sparse observations, and transforms into a generative synthesis engine when leveraging the DiT-based map decoder to reconstruct fields from radio environmental priors.

\begin{figure*}[!t]
\centering
\includegraphics[width=\textwidth]{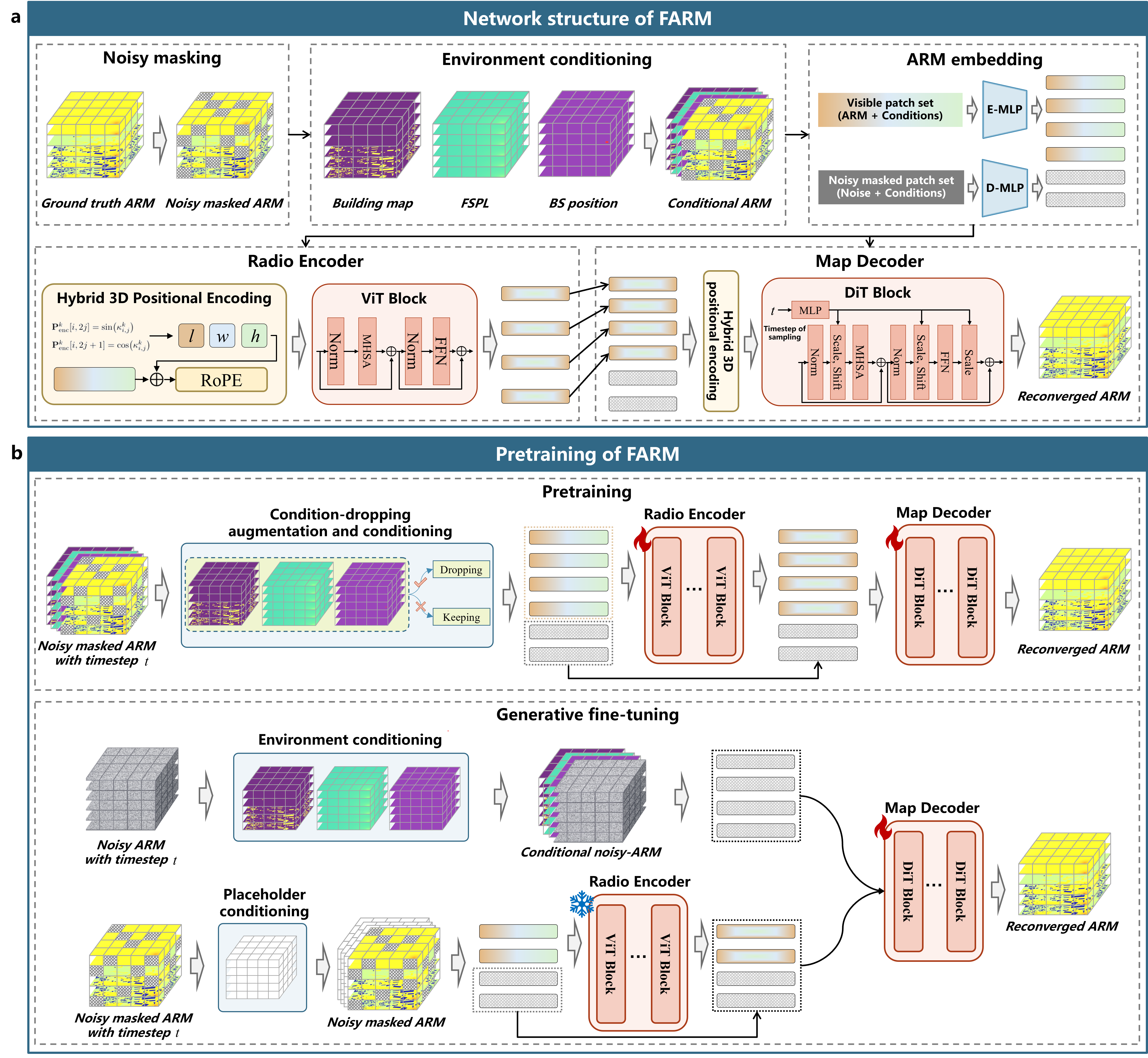}
\caption{\textbf{Overview of FARM:} \textbf{a.} Architecture of FARM integrating a MAE module with a diffusion module. \textbf{b.} Two-stage training strategy comprising self-supervised pretraining and generative fine-tuning.}
\label{fig:Overview_of_FARM}
\end{figure*}

\subsection{Network Structure}

As illustrated in Fig.\ref{fig:Overview_of_FARM}, the FARM framework comprises five functional modules: noisy masking, environment conditioning, ARM embedding, radio encoder, and map decoder. Each module is detailed below.

\subsubsection{\textbf{Noise Masking}}

To align the representation learning of the encoder with the generative capabilities of the decoder, FARM performs masking in the voxel space rather than the latent space. The ground-truth ARM $\mathbf{R}$ is first partitioned into a sequence of non-overlapping 3D patches of size $(l_p, w_p, h_p)$. This partitioning yields a total of $N_p = \frac{L}{l_p} \times \frac{W}{w_p} \times \frac{H}{h_p}$ patches. The patch-level noisy masking operation is then applied as follows:
\begin{equation}
    \mathbf{R}^{\text{mask}} = \mathrm{Mask}(\mathbf{R} \mid p_{\mathrm{mask}}, t),
\end{equation}
where $\mathrm{Mask}(\cdot)$ denotes a stochastic masking operator parameterized by the masking ratio $p_{\mathrm{mask}} \in [0, 1]$ and the flow-matching time step $t \in [0, 1]$ defined in Eq. \eqref{eq5}. Under this operation, $N_{m} = \lfloor p_{\text{mask}} N_p \rfloor$ patches are selected for masking. Crucially, for these $N_m$ selected patches, the original signal is replaced by the noisy intermediate state $\mathbf{Z}_t$ from the flow-matching probability path. The remaining $N_v = N_p - N_m$ visible patches are preserved to provide the encoder with structural context. This joint corruption strategy ensures that the model simultaneously learns to interpolate missing spatial structures (MAE objective) and denoise Gaussian samples into high-fidelity radio distributions, i.e., the diffusion objective.

\subsubsection{\textbf{Environment Conditioning}}

To incorporate heterogeneous coverage dimensions, carrier frequencies, antenna parameters, and blockage distributions into the FARM framework, we construct three 3D voxel grids that share the spatial coordinate system of the input ARM. These grids represent the BS position, free-space pathloss (FSPL), and building occupancy, providing spatially aligned physical priors. The BS position $\mathbf{p}_{tx}$ defines the geometric relationship between the transmitter and each spatial voxel. We represent this positional cue as a binary voxel grid $\mathbf{V}^{\mathrm{pos}} \in \{0, 1\}^{L \times W \times H}$, where only the voxel corresponding to the transmitter coordinates $(l_{tx}, w_{tx}, h_{tx})$ is activated:
\begin{equation}
\mathbf{V}^{\mathrm{pos}}_i=
\begin{cases}
1, & (l_i,w_i,h_i)=(l_{tx},w_{tx},h_{tx}),\\
0, & \text{otherwise}. 
\end{cases} \quad i \in \mathcal{V}
\end{equation}
The BS configuration, including the carrier frequency $f_c$ and antenna parameters $\mathbf{a}$, governs large-scale propagation characteristics. We parameterize this information as an FSPL voxel grid $\mathbf{V}^{\mathrm{fspl}} \in \mathbb{R}^{L \times W \times H}$ to provide a spatially aligned physical anchor. For the $i$-th voxel, the distance to the transmitter is $d_i = \| \mathbf{x}_i - \mathbf{p}_{tx} \|_2 \cdot \Delta$, where $\Delta$ denotes the spatial resolution. The FSPL value at each voxel is then computed as:
\begin{equation}
    \mathbf{V}^{\mathrm{fspl}}_i = 20\log_{10}(d_{i}) + 20\log_{10}(f_c) - g_{i}(\mathbf{a}) + C,
\end{equation}
where $g_{i}(\mathbf{a})$ is the antenna gain toward the $i$-th voxel and $C$ is a constant determined by system units. As a complementary structural prior, the building occupancy grid $\mathbf{B}$ encodes blockage constraints. Since $\mathbf{B}$ is inherently aligned with the spatial characteristics of the ARM, it is integrated directly. These grids are concatenated with the noise-masked ARM to form the conditioned input tensor:
\begin{equation}
    \label{eq13}
    \mathbf{R}^{\mathrm{cond}} = \mathbf{R}^{\text{mask}} \oplus \mathbf{V}^{\mathrm{pos}} \oplus \mathbf{V}^{\mathrm{fspl}} \oplus \mathbf{B} \in \mathbb{R}^{4 \times L \times W \times H}.
\end{equation}
The resulting $\mathbf{R}^{\mathrm{cond}}$ is subsequently partitioned into a visible patch set $\mathcal{P}^{\mathrm{vis}} = \{ \mathbf{R}_1^p, \dots, \mathbf{R}_{N_v}^p \}$ for representation learning and a noise-masked patch set $\mathcal{P}^{\mathrm{noi}} = \{ \tilde{\mathbf{R}}_1^p, \dots, \tilde{\mathbf{R}}_{N_m}^p \}$ for generative denoising. Compared to cross-attention conditioning methods, our channel-wise concatenation strategy avoids the heavy computational overhead associated with latent extraction and attention score calculation.

\subsubsection{\textbf{ARM Embedding}}

To facilitate network processing, the 3D patches are converted into 1D sequential tokens compatible with the Transformer architecture. Each visible patch and noise-masked patch is first flattened into a high-dimensional vector. Because the visible patches in $\mathcal{P}^{\mathrm{vis}}$ provide clean contextual information for the encoder, while the noisy masked patches in $\mathcal{P}^{\mathrm{noi}}$ serve as corrupted reconstruction targets for the decoder, FARM employs two distinct linear projection layers (MLPs) to map these vectors into their respective feature spaces. Specifically, encoder-side and decoder-side MLPs project these vectors into feature tokens as follows:
\begin{equation}
\begin{aligned}
\mathbf{f}^{\mathrm{vis}}_i &= \mathrm{E\text{-}MLP}(\mathbf{R}^p_i), \quad \mathbf{R}^p_i \in \mathcal{P}^{\mathrm{vis}},\\
\mathbf{f}^{\mathrm{noi}}_i &= \mathrm{D\text{-}MLP}(\tilde{\mathbf{R}}^p_i), \quad \tilde{\mathbf{R}}^p_i \in \mathcal{P}^{\mathrm{noi}}.
\end{aligned}
\end{equation}
The $\mathrm{E\text{-}MLP}(\cdot)$ operator maps the flattened visible patches of dimension $4 l_p w_p h_p$ to the encoder hidden dimension $D_{\mathrm{enc}}$. Conversely, the $\mathrm{D\text{-}MLP}(\cdot)$ operator projects the noisy counterparts into the decoder input dimension $D_{\mathrm{dec}}$. The resulting tokens are aggregated into the feature matrices $\mathbf{F}^{\mathrm{vis}} \in \mathbb{R}^{N_v \times D_{\mathrm{enc}}}$ and $\mathbf{F}^{\mathrm{noi}} \in \mathbb{R}^{N_m \times D_{\mathrm{dec}}}$. To preserve the spatial topology of the aerial radio environment, 3D positional encodings are added to these tokens. This ensures the model retains awareness of the relative coordinates of each patch within the volumetric grid. This dual-path embedding strategy allows FARM to maintain a sharp distinction between observed environmental cues and the generation targets.

\subsubsection{\textbf{Radio Encoder}}

A ViT-based radio encoder is designed to transform visible tokens into compact, high-level representations of the ARM. To accurately reflect the complex spatial characteristics of the radio environment, this encoder is equipped with a hybrid 3D positional encoding (H-PE) scheme. This scheme utilizes absolute positional encoding to capture the global spatial structure while employing relative positional encoding to model the underlying propagation features.

For the first component of the H-PE, the absolute positional encoding explicitly injects 3D spatial priors into the visible tokens \cite{Vasw_Nips_2017}. To accurately reflect the spatial structure, the embedding dimension $D_{\mathrm{enc}}$ is partitioned into three axis-specific subspaces satisfying $D_{\mathrm{enc}} = D_l + D_w + D_h$, defined as $D_l = D_w = \lfloor D_{\mathrm{enc}}/3 \rfloor$, and $\quad D_h = D_{\mathrm{enc}} - 2\lfloor D_{\mathrm{enc}}/3 \rfloor$. To ensure robust generalization across varying region sizes, a SinCos encoding is adopted for each spatial axis. For the $i$-th visible token, the angle encoding for coordinate $p_i^k$ in any given axis $k \in \{l, w, h\}$ is calculated as $\kappa_{i,j}^k=\frac{p_i^k}{10000^{2j/D_k}}$, where $j \in [0, D_k/2)$ is the dimension index. Thus, the corresponding absolute positional encoding is denoted as:
\begin{equation}
\mathbf{P}^{\mathrm{enc},k}[i, 2j] = \sin\!\left(\kappa_{i,j}^k\right), \quad
\mathbf{P}^{\mathrm{enc},k}[i, 2j+1] = \cos\!\left(\kappa_{i,j}^k\right),
\end{equation}
Finally, these independent components are concatenated along the feature dimension to form the complete 3D absolute positional encoding $\mathbf{P}^{\mathrm{enc}}$, with $\mathbf{P}^{\mathrm{enc}}[i,:]=\mathbf{P}^{\mathrm{enc},l}[i,:]\oplus\mathbf{P}^{\mathrm{enc},w}[i,:]\oplus\mathbf{P}^{\mathrm{enc},h}[i,:]$.

To dynamically model the propagation features, the H-PE additionally employs rotary position embedding (RoPE) to capture the distance-dependent attenuation for signal propagation. This approach embeds 3D positions through coordinate-dependent rotations directly into the attention-score calculation \cite{SLPM_Neurocomputing_2024}. To match the pairwise rotation structure of RoPE, each angle encoding is duplicated to form an expanded angle vector $\boldsymbol{\kappa}_{i}^{k} \in \mathbb{R}^{D_k}$, such that $\boldsymbol{\kappa}_{i}^{k}[2j] = \boldsymbol{\kappa}_{i}^{k}[2j+1] = \kappa_{i,j}^{k}$. The full rotation phase is obtained as $\boldsymbol{\kappa}_i = [\boldsymbol{\kappa}_i^{l}; \boldsymbol{\kappa}_i^{w}; \boldsymbol{\kappa}_i^{h}] \in \mathbb{R}^{1 \times D_{\mathrm{enc}}}$. For each Transformer block, let $\mathbf{Q}, \mathbf{K} \in \mathbb{R}^{N_v \times D_{\mathrm{enc}}}$ denote the query and key matrices. 
The RoPE is applied as:
\begin{equation}
\label{eq17}
\begin{aligned}
\widetilde{\mathbf{Q}}[i,:] &= \mathbf{Q}[i,:] \odot \cos(\boldsymbol{\kappa}_i) + \mathrm{rot}\!\left(\mathbf{Q}[i,:]\right) \odot \sin(\boldsymbol{\kappa}_i), \\
\widetilde{\mathbf{K}}[i,:] &= \mathbf{K}[i,:] \odot \cos(\boldsymbol{\kappa}_i) + \mathrm{rot}\!\left(\mathbf{K}[i,:]\right) \odot \sin(\boldsymbol{\kappa}_i).
\end{aligned}
\end{equation}
Here, $\mathrm{rot}(\cdot)$ is the 2D rotation function, which rearranges adjacent feature pairs such that, for an input vector $\mathbf{x}=[x_0, x_1, x_2, x_3, \dots]$, $\mathrm{rot}(\mathbf{x})=[-x_1, x_0, -x_3, x_2, \dots]$. By evaluating attention scores using these rotated matrices, the relative distance penalty is naturally applied, simulating the spatial attenuation of radio propagation. Finally, letting $\tilde{\mathcal{G}}^{\mathrm{enc}}(\cdot)$ represent the operation of the transformer blocks equipped with RoPE, the encoder output $\mathbf{F}^{\mathrm{enc}} \in \mathbb{R}^{N_v \times D_{\mathrm{enc}}}$ is derived as:
\begin{equation}
\mathbf{F}^{\mathrm{enc}}=\tilde{\mathcal{G}}^{\mathrm{enc}}(\mathbf{F}^{\mathrm{vis}}+\mathbf{P}^{\mathrm{enc}}).
\end{equation}

\subsubsection{\textbf{Map Decoder}}

The map decoder is designed to execute the reverse diffusion process in the voxel space to reconstruct the high-fidelity ARM. Unlike the patch recovery decoder found in standard MAE architectures, the objective of this module is timestep-conditional denoising. At each diffusion step $t$, a sequence of Transformer blocks iteratively updates the output based on the clean visible context extracted by the radio encoder. To facilitate this, $\mathbf{F}^{\mathrm{enc}}$ is first converted to $\widetilde{\mathbf{F}}^{\mathrm{enc}}$ via an MLP to align with the decoder feature dimension $D_{\mathrm{dec}}$. Subsequently, $\widetilde{\mathbf{F}}^{\mathrm{enc}}$ and the noisy masked tokens $\mathbf{F}^{\mathrm{noi}}$ are concatenated to form the initial decoder sequence $\mathbf{F}^{\mathrm{full}} \in \mathbb{R}^{N_p \times D_{\mathrm{dec}}}$. The spatial topology is maintained by adding the decoder positional encoding $\mathbf{P}^{\mathrm{dec}} \in \mathbb{R}^{N_p \times D_{\mathrm{dec}}}$ and applying RoPE within the decoder transformer blocks.

The timestep embedding maps the diffusion timestep $t$ to a conditioning vector $\mathbf{t}^{\mathrm{dec}} \in \mathbb{R}^{1 \times d_t}$ with dimension $d_t$ to control the denoising behavior. Specifically, we compute $\mathbf{t}^{\mathrm{dec}} = \mathrm{FFN}(\mathrm{SiLU}(\mathrm{FFN}(\mathbf{e}_t)))$, where $\mathrm{FFN}(\cdot)$ denotes a feed-forward network and $\mathbf{e}_t = [\cos(t\boldsymbol{\omega}), \sin(t\boldsymbol{\omega})]$ represents a sinusoidal encoding with frequency vector $\boldsymbol{\omega}$, which provides a smooth and continuous representation across diffusion stages. For the $b$-th transformer block, this vector is utilized to generate six modulation parameters via an MLP, following the AdaLN-Zero paradigm:
\begin{equation}
[\boldsymbol{\beta}_{1,b},\boldsymbol{\gamma}_{1,b},\boldsymbol{\alpha}_{1,b},
\boldsymbol{\beta}_{2,b},\boldsymbol{\gamma}_{2,b},\boldsymbol{\alpha}_{2,b}] = \mathrm{MLP}(\mathbf{t}^{\mathrm{dec}}).
\end{equation}
Here, $\boldsymbol{\beta}$ and $\boldsymbol{\gamma}$ denote the shift and scale vectors for the adaptive normalization, while $\boldsymbol{\alpha}$ represents the gating parameters for the residual paths.

The input sequence $\mathbf{F}^{\mathrm{full}\text{-}\mathrm{pe}} = \mathbf{F}^{\mathrm{full}} + \mathbf{P}^{\mathrm{dec}}$ undergoes adaptive layer normalization using these condition-derived parameters:
\begin{equation}
\mathbf{F}^{\mathrm{in}}_b[i,:]
=
(1+\boldsymbol{\gamma}_{1,b})\odot
\frac{\mathbf{F}^{\mathrm{full}\text{-}\mathrm{pe}}_{b-1}[i,:]}
{\sqrt{\frac{1}{D_{\mathrm{dec}}}\sum_{j=1}^{D_{\mathrm{dec}}}\left(\mathbf{F}^{\mathrm{full}\text{-}\mathrm{pe}}_{b-1}[i,j]\right)^2}}
+\boldsymbol{\beta}_{1,b},
\end{equation}
Let $\mathbf{W}^Q_{b}, \mathbf{W}^K_{b}, \mathbf{W}^V_{b} \in \mathbb{R}^{D_{\mathrm{dec}} \times D_{\mathrm{dec}}}$ denote the query, key, and value projection matrices in the $b$-th transformer block of the decoder. The query, key, and value matrices are obtained by $\mathbf{Q}_b=\mathbf{F}^{\text{in}}_b\mathbf{W}^Q_{b}$, $\mathbf{K}_b=\mathbf{F}^{\text{in}}_b\mathbf{W}^K_{b}$ and $\mathbf{V}_b=\mathbf{F}^{\text{in}}_b\mathbf{W}^V_{b}$. According to Eq.~\eqref{eq17}, the next self-attention module with RoPE is expressed as:
\begin{equation}
\mathbf{F}^{\mathrm{attn}}_b = \operatorname{Softmax} \left( \frac{\widetilde{\mathbf{Q}}_b (\widetilde{\mathbf{K}}_b)^{\text{T}}}{\sqrt{D_{\mathrm{dec}}}} \right) \mathbf{V}_b.
\end{equation} 
The attention output is gated using the condition-derived parameter as $\mathbf{F}^{\mathrm{gate}}_b = \boldsymbol{\alpha}_{1,b}\odot \mathbf{F}^{\mathrm{attn}}_b$. After a second stage of adaptive normalization and feed-forward processing using $\boldsymbol{\beta}_{2,b}$, $\boldsymbol{\gamma}_{2,b}$, and $\boldsymbol{\alpha}_{2,b}$, the final block output $\mathbf{F}^\mathrm{out}_b$ is obtained. The refined latent features from the terminal block are projected back into the voxel space via an MLP to produce the final reconstructed ARM $\hat{\mathbf{R}} \in \mathbb{R}^{L \times W \times H}$. This output supports both patch-wise reconstruction for representation learning and conditional generative modeling for map synthesis. Letting $\tilde{\mathcal{G}}^{\mathrm{dec}}(\cdot)$ denote the decoder blocks equipped with RoPE, the reconstruction operation is expressed as:
\begin{equation}
\hat{\mathbf{R}}=\tilde{\mathcal{G}}^{\mathrm{dec}}(\mathbf{F}^{\mathrm{full}}+\mathbf{P}^{\mathrm{dec}} \,|\, t).
\end{equation}

\subsection{Model Training and Inference}

To unify condition-based and condition-free ARM construction, FARM adopts a two-stage training strategy comprising self-supervised pretraining and generative fine-tuning. During the pretraining stage, all parameters are optimized using a random masking strategy to learn fundamental representations of the aerial radio environment. A condition-dropping augmentation is introduced in this phase to prevent the model from becoming overly dependent on radio environmental priors. This strategy enhances the robustness of the framework, ensuring high performance in both condition-based and condition-free construction paradigms.

Because the pretraining stage focuses primarily on broad representation learning, a generative fine-tuning stage is subsequently introduced to enhance high-fidelity construction capabilities. Unlike existing methods that are typically restricted to specific datasets, FARM is pretrained on heterogeneous data across various domains. This exposure allows the model to generalize effectively to unseen spatial geometries and radio configurations. During inference, FARM seamlessly adapts to different input cases of condition-based and condition-free construction paradigms by flexibly coordinating the radio encoder and map decoder. Specifically, FARM operates in a decoder-only mode for the condition-only input case and in a full encoder-decoder mode for condition-free/condition-and-sample input cases. This versatile pipeline ensures the framework can serve as a reliable foundation for diverse sensing and communication tasks in the low-altitude economy.

\subsubsection{\textbf{Self-supervised Pretraining}}

During the pretraining stage, the radio encoder and the map decoder are jointly optimized using masked denoising with velocity-space supervision. For each training iteration, the input is partially corrupted by a noise sample at a random timestep $t$ and concatenated with the three environmental voxel grids for conditioning. To ensure that FARM can transition seamlessly between condition-based and condition-free construction, we introduce a condition-dropping augmentation strategy. This approach simulates the absence of radio environmental priors by assigning a specific null value to the condition channels. Since the input is normalized to the range $[-1, 1]$ to improve optimization stability, we utilize a value of $-2$ to represent missing information. This value lies outside the effective data distribution, allowing the model to clearly distinguish between valid physical priors and their absence. Specifically, let $m \sim \mathrm{Bernoulli}(p_m)$ denote a condition-dropping indicator where $m=1$ signifies that conditions are dropped and $m=0$ indicates they are retained. We define a constant voxel grid $\mathbf{M} \in \mathbb{R}^{L \times W \times H}$ with all entries set to $-2$. The conditional channels are then processed as follows:
\begin{equation}
\label{eq22}
\begin{aligned}
\widetilde{\mathbf{V}}^\mathrm{fspl} &= (1-m)\mathbf{V}^\mathrm{fspl}+m\mathbf{M}, \\
\widetilde{\mathbf{V}}^\mathrm{pos} &= (1-m)\mathbf{V}^\mathrm{pos}+m\mathbf{M}, \\
\widetilde{\mathbf{B}} &= (1-m)\mathbf{B}+m\mathbf{M}. 
\end{aligned}
\end{equation}
The resulting augmented input is expressed as $\widetilde{\mathbf{R}}^\mathrm{cond} = \mathbf{R}^{\text{mask}} \oplus \widetilde{\mathbf{V}}^\mathrm{pos} \oplus \widetilde{\mathbf{V}}^\mathrm{fspl} \oplus \widetilde{\mathbf{B}}$. By exposing the framework to both condition-present and condition-absent inputs during the pretraining phase, FARM learns a unified representation that is effective across the entire spectrum of ARM construction tasks.

Subsequently, the visible and noise-masked patch sets are embedded into tokens and processed by the radio encoder and map decoder, respectively. Upon obtaining the predicted ARM, a velocity-space loss function is employed to optimize the complete set of model parameters $\mathbf{\Theta}$. Computed exclusively over the masked patches, this loss ensures the model learns to reconstruct the missing volumetric data while adhering to the probability path of the flow-matching objective. The objective function is defined as:
\begin{equation}
\label{eq23}
\mathcal{L}(\mathbf{\Theta}) \!=\! \mathbb{E}_{t, \mathbf{R}^{p}, \boldsymbol{\epsilon}} \!\!\left[\!\frac{1}{N_m}\!\sum_{i =1}^{N_m} \left\| \! \left( \!\frac{\hat{\mathbf{R}}^{p}_{i} \!- \!\mathbf{R}^p_{t,i}}{1 - t} \!\!\right) \! - \!\left( \mathbf{R}^{p}_{i} \!-\! \boldsymbol{\epsilon} \right) \!\right\|_2^2 \right],
\end{equation}
where $\mathbf{R}^p_{t,i} = t\mathbf{R}^p_{i} + (1-t)\boldsymbol{\epsilon}$ represents the $i$-th noisy intermediate patch in the masked set $\mathcal{P}^{\mathrm{noi}}$ and $\mathbf{R}^{p}_{i}$ denotes the corresponding ground-truth patch. By minimizing the discrepancy between the predicted and target velocities in the voxel space, the pretraining phase enables the framework to capture the complex spatial correlations and global structures of the aerial radio environment. This foundation allows for robust generalization across heterogeneous BS configurations.

\subsubsection{\textbf{Map Decoder Fine-tuning}}

Because the map decoder is not fully optimized during pretraining, we further fine-tune it solely to improve the ARM construction ability of FARM. For each batch, two objectives corresponding to condition-based and condition-free tasks are optimized sequentially. For the condition-based construction branch, the condition channels are kept, yielding $\widetilde{\mathbf{R}}^{\mathrm{cond}}_{\mathrm{based}}=\mathbf{R}^{\text{mask}} \oplus \widetilde{\mathbf{V}}^\mathrm{pos} \oplus \widetilde{\mathbf{V}}^\mathrm{fspl} \oplus \widetilde{\mathbf{B}}$. After patchification and timestep sampling, full masking is applied to the radio patches by setting $p_{\mathrm{mask}}=1$, which results in $N_v=0$ visible radio tokens. Therefore, the map decoder performs condition-guided full-map construction and the loss is computed as $\mathcal{L}_{\mathrm{based}}(\mathbf{\Theta}^{\mathrm{dec}})$ using Eq.~\eqref{eq23}. For the condition-free construction branch, all condition channels are replaced by the missing-condition sentinel value, i.e.,
$\widetilde{\mathbf{R}}^{\mathrm{cond}}_{\mathrm{free}} = \mathbf{R}\oplus \mathbf{M}\oplus \mathbf{M}\oplus \mathbf{M}$.
The masking ratio $p_{\mathrm{mask}}<1$ is adopted so that the map decoder reconstructs the masked radio patches based on the visible radio context. The corresponding loss is calculated by $\mathcal{L}_{\mathrm{free}}(\mathbf{\Theta}^{\mathrm{dec}})$ based on Eq.~\eqref{eq23}. After finishing both forward propagations, the fine-tuning loss function is computed as:
\begin{equation}
\mathcal{L}_{\mathrm{ft}}(\mathbf{\Theta}^{\mathrm{dec}})
=
\lambda_{\mathrm{free}}\mathcal{L}_{\mathrm{free}}(\mathbf{\Theta}^{\mathrm{dec}})
+
\lambda_{\mathrm{based}}\mathcal{L}_{\mathrm{based}}(\mathbf{\Theta}^{\mathrm{dec}}),
\end{equation}
where $\lambda_{\mathrm{free}}$ and $\lambda_{\mathrm{based}}$ represent the weights values. Subsequently, this overall loss function is utilized to calculate the gradients for the map decoder parameters update.

\subsubsection{\textbf{Inference}}

During the inference phase, FARM supports three distinct operational settings to accommodate different levels of available information. In all scenarios, the reverse diffusion process originates from standard Gaussian white noise rather than an interpolated state. The masked radio patches are initialized as $\mathbf{Z}_0 \sim \mathcal{N}(\mathbf{0}, \mathbf{I})$. At each denoising step, the map decoder predicts the clean ARM in the voxel space, and this prediction is converted into a velocity vector to update the intermediate state through the ODE solver until the final timestep is reached.

For the condition-free task, FARM operates in an encoder-decoder mode with partial masking. The three conditional channels are set to the null grid $\mathbf{M}$, and the sparse observed radio patches are utilized as the visible context. The radio encoder extracts latent representations from these sparse observations to guide the map decoder in reconstructing the complete field. For the condition-based task, FARM transitions to a decoder-only configuration with full masking for the condition-only input case. In this mode, the noise-initialized ARM and the three validated conditional grids are fed directly into the map decoder. The complete ARM is constructed solely through condition-guided denoising, leveraging the learned physical priors to map environmental features to signal distributions. For the condition-and-sample input case, which involves both radio environmental priors and physical signal samples, FARM utilizes the full encoder-decoder pipeline. The visible ARM patches and the three conditional channels are embedded into tokens and processed by the radio encoder to produce integrated radio-environment representations. Simultaneously, the noise-initialized masked patches and the same conditional channels are embedded into noisy tokens. The map decoder then processes the concatenation of these two token sets to reconstruct the high-resolution ARM. This flexible execution logic allows FARM to maximize the utility of all available data sources in the low-altitude economy.

\section{ARM-Omni Dataset for FARM}
\label{sec:dataset}

ARM-Omni is a large-scale aerial radio environment dataset constructed using the TensorFlow-based ray tracing simulator Sionna RT. We design this dataset to address two critical limitations identified in existing public repositories: the restriction to narrow receiver height ranges that exclude the majority of the low-altitude operating envelope, and the absence of joint multi-band and multi-antenna diversity required for configuration-aware learning. The following sections first detail the dataset generation pipeline, and then describe the various transmission configurations and comprehensive height-coverage design that characterize ARM-Omni.

\subsection{Dataset Generation Pipeline}

\begin{figure}[!t]
\centering
\includegraphics[width=\columnwidth]{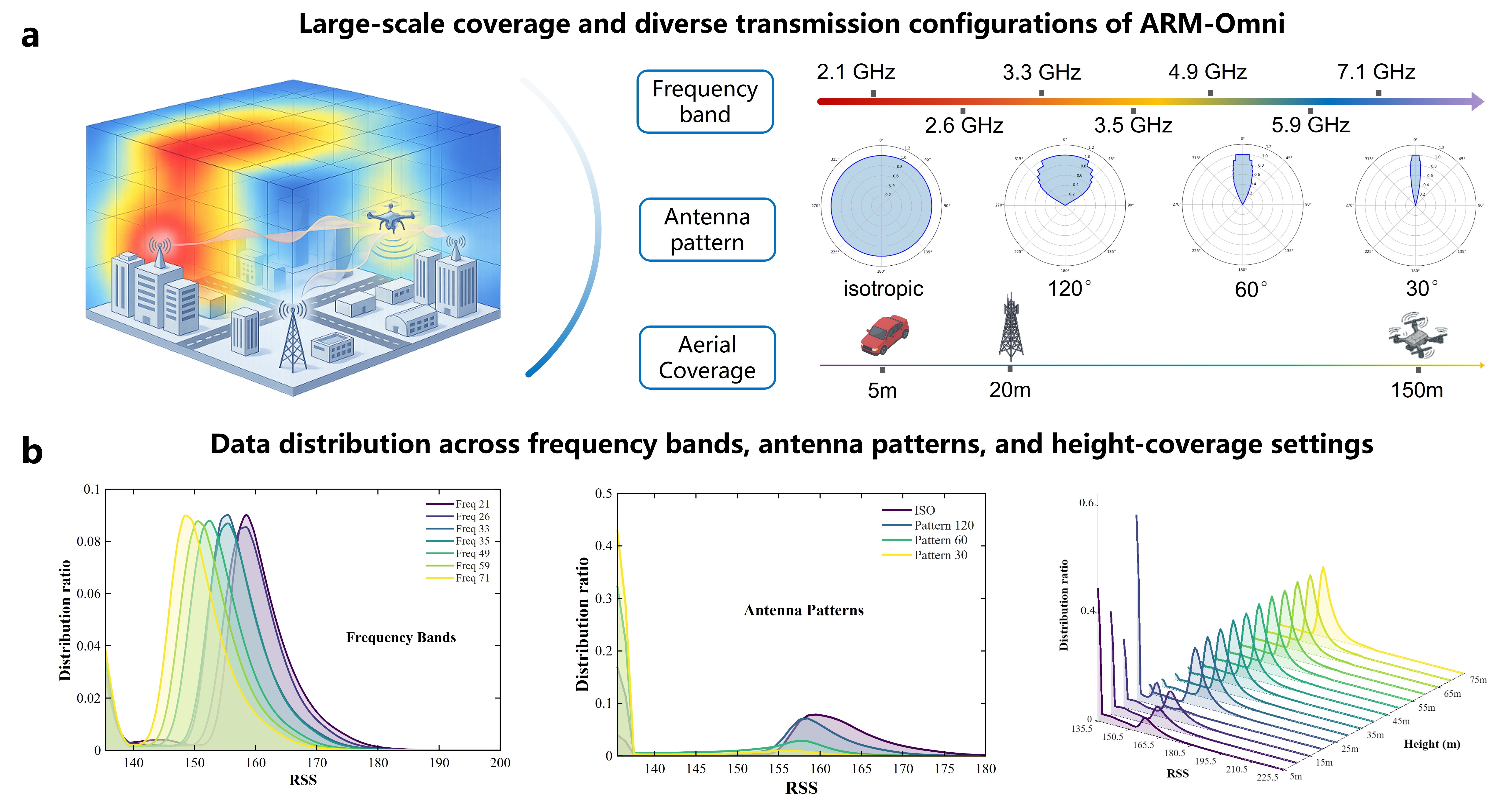}
\caption{\textbf{\textbf{Dataset illustration:}} \textbf{a.} ARM-Omni features diverse transmission configurations and a large-scale height coverage. \textbf{b.} Distribution analysis of ARM across frequency bands on the isotropic antenna (left); across antenna patterns at 2.1 GHz (middle); and across slices on the isotropic antenna at 2.1 GHz (right). The distributions are consistent with expected propagation trends: Higher frequencies yield stronger attenuation, reducing the proportion of high-RSS regions; narrower beams localize coverage, expanding no-signal regions; and higher receiver slices reduce shadowing induced by blockage, thereby leading to a more Gaussian-like distribution.}
\label{fig:dataset}
\end{figure}

The construction of ARM-Omni follows a rigorous three-stage pipeline designed around the previously defined radio-environmental priors, including environmental structure modeling, BS location selection, and ray tracing under diverse transmission configurations. Specifically, we first collect $130$ diverse urban scenes from OpenStreetMap, covering residential, commercial, and mixed-use districts across different regions, and use the corresponding building footprints and elevation data to construct a physically accurate 3D urban environment. Second, for each $1000 \text{m} \times 1000 \text{m}$ area, a high-resolution height map is generated with a $1\text{m} \times 1\text{m}$ resolution. This map serves as an environmental prior for BS location selection, filtering out geometrically constrained or heavily occluded candidates while favoring positions with higher line-of-sight (LoS) probability. $100$ transmitter positions per scenario are sampled using a grid-based strategy to ensure spatial diversity. Finally, given the constructed 3D environments and selected BS locations, Sionna RT performs ray tracing under diverse transmission configurations, as illustrated in Fig. \ref{fig:dataset}a. To better approximate real-world propagation, reflections, diffractions, and material-specific propagation properties are considered when computing the RSS at each receiver voxel. The complete pipeline generates approximately $26.4$ million radio maps, providing the scale necessary for foundation model pretraining across heterogeneous BS settings.

\subsection{Multi-band and Multi-antenna Configurations}
\label{sec:dataset:rf}

As illustrated in Fig. \ref{fig:dataset}a, our simulations span seven carrier frequencies essential to modern and future networking \cite{Rappaport_npjWT_2025}: 2.1 GHz (AWS), 2.6 GHz (LTE B7), 3.3 GHz and 3.5 GHz (C-band), 4.9 GHz (5G-NR n79), 5.9 GHz (ITS band), and 7.1 GHz (5G upper mid-band). This frequency diversity captures distinct propagation behaviors, including frequency-dependent free-space path-loss scaling and material penetration characteristics. Such diversity allows FARM to learn frequency-dependent propagation features, facilitating cross-band generalization. Additionally, four common cellular radiation patterns \cite{Banerjee_npjWT_2026}, ranging from narrow-beam to wide-sector coverage, are incorporated: an isotropic pattern and three directional antenna patterns with half-power beamwidths of $30^\circ$, $60^\circ$, and $120^\circ$. This antenna-pattern diversity exposes FARM to different spatial radiation characteristics, enabling it to model coverage anisotropy and improve cross-antenna generalization. To capture spatial asymmetry, we further sample three random yaw angles per BS position. This multi-band and multi-antenna design enriches ARM-Omni with diverse propagation conditions under a unified simulation framework and encourages FARM to learn deep radio representations for cross-scenario generalization improvement.

\subsection{Aerial Height Coverage}
\label{sec:dataset:height}
A defining feature of the aerial radio environment is the vertical heterogeneity of signal propagation. As shown in Fig. \ref{fig:dataset}b, near-ground radio maps exhibit strong spatial variability due to blockage-induced shadowing and multipath propagation. As the receiver height increases, blockage effects are progressively reduced and LoS components become more prevalent, leading to smoother RSS distributions. However, existing datasets fail to capture this transition. As shown in Table~\ref{tab:dataset_comparison}, the most extensive prior work covers up to $20\text{m}$, leaving the upper low-altitude operating range largely unrepresented. To ensure broad coverage of the low-altitude operating envelope, we configure ARM-Omni to span a vertical range from $5\text{m}$ to $150\text{m}$ across $30$ uniformly spaced height levels with a $5\text{m}$ vertical interval. This design substantially extends the vertical coverage while preserving a manageable volumetric data size, enabling ARM-Omni to characterize the transition from blockage-dominated near-ground propagation to LoS-dominant aerial propagation. This design also allows FARM to jointly model the vertical evolution of radio propagation within a single 3D input, rather than treating heights as independent slices. This approach enables the model to exploit statistical dependencies between layers, which is particularly vital for condition-free construction from sparse vertical observations. Furthermore, the extensive height coverage facilitates altitude extrapolation. Together, these features provide the necessary framework for a critical digital twin of the low-altitude economy.

\section{Experiments}
\label{sec:exp}

\subsection{Experiment Setup}

All experiments are conducted using the proposed ARM-Omni dataset. The total of $130$ scenarios is evenly partitioned into $13$ task-oriented subsets, as summarized in Table~\ref{tab:exp_datasets}. Subsets D1--D10 serve as the in-domain datasets for both model training and standard evaluation. Within each of these subsets, samples are randomly split into training, validation, and testing sets using an 8:1:1 ratio. The remaining three subsets are reserved for zero-shot out-of-domain evaluation under unseen configurations. Specifically, P1 extends the maximum receiver height from $130m$ to $150m$ and narrows the coverage range to one-fourth of the original to assess spatial transferability. F1 introduces unseen carrier frequencies at 2.6 GHz and 7.1 GHz to verify cross-frequency generalization, and A1 incorporates an unseen antenna radiation pattern with a $60^\circ$ beamwidth to evaluate pattern adaptation. The training and evaluation datasets are available at \nolinkurl{https://huggingface.co/datasets/jliang097/FARM_training_test}.

For the proposed FARM, we evaluate the three model sizes listed in Table~\ref{tab:fmure_sizes}, adopting FARM-base as the default architecture. For the input ARM, the patch size is set to $(l_p, w_p, h_p)=(32, 32, 2)$. During the pretraining stage, the condition-dropping probability is maintained at $p_m=0.2$ and the masking ratio is fixed at $p_{\mathrm{mask}}=75\%$. During the fine-tuning stage, full masking is applied for condition-based construction, while $p_{\mathrm{mask}}=95\%$ is employed for condition-free construction. The corresponding losses are combined equally with $\lambda_{\mathrm{free}}=\lambda_{\mathrm{based}}=1$. All training procedures are performed on a server equipped with four NVIDIA GeForce A800 GPUs using mixed-precision training. The model is trained with a global batch size of 100. Parameters are optimized using the AdamW optimizer \cite{LoHu_ICLR_2019} with momentum hyperparameters $\beta_1=0.9$ and $\beta_2=0.999$. For the two-stage training strategy, FARM is first pretrained for $80$ epochs with a peak learning rate of $2 \times 10^{-4}$, followed by a fine-tuning stage of $20$ epochs at a reduced learning rate of $1 \times 10^{-4}$. For training stability, a 5-epoch linear warmup is used during the pretraining stage. The number of inference steps is set to $2$ for zero-shot experiments on P1 and $1$ for all other evaluation scenarios. Unless otherwise specified, the sampling region is set to the size of one patch and the sampling rate is fixed at $5\%$ across all experiments.

\begin{table}[t]
\centering
\caption{Constructed datasets based on ARM-Omni for training and evaluation.}
\label{tab:exp_datasets}
\tiny
{\setlength{\tabcolsep}{1pt}
\renewcommand{\arraystretch}{1.5}
\begin{tabular}{>{\centering\arraybackslash}p{0.075\columnwidth}>{\centering\arraybackslash}p{0.22\columnwidth}>{\centering\arraybackslash}p{0.155\columnwidth}>{\centering\arraybackslash}p{0.17\columnwidth}>{\centering\arraybackslash}p{0.19\columnwidth}>{\centering\arraybackslash}p{0.09\columnwidth}@{}}
\toprule
\textbf{Dataset} & \multicolumn{1}{c}{\textbf{Frequencies (GHz)}} & \multicolumn{1}{c}{\textbf{Max Rx Height (m)}} & \textbf{Beamwidths} & \multicolumn{1}{c}{\textbf{Map Grid Size }} & \textbf{Volume} \\
\midrule
D1  & 2.1, 3.3, 5.9 & 120  & $30^\circ, 120^\circ$, Iso & 512 $\times$ 512 & 15000 \\
D2  & 3.5, 4.9, 5.9 & 130  & $30^\circ, 120^\circ$, Iso & 512 $\times$ 512 & 15000 \\
D3  & 2.1, 4.9, 5.9 & 110  & $30^\circ, 120^\circ$ & 512 $\times$ 512 & 15000 \\
D4  & 3.3, 3.5, 4.9 & 110  & $30^\circ, 120^\circ$, Iso & 512 $\times$ 512 & 15000 \\
D5  & 3.3, 4.9, 5.9 & 120  & $30^\circ, 120^\circ$, Iso & 512 $\times$ 512 & 15000 \\
D6  & 2.1, 3.3, 3.5 & 130  & $30^\circ, 120^\circ$ & 512 $\times$ 512 & 15000 \\
D7  & 2.1, 3.3, 3.5, 5.9 & 130  & $30^\circ$, Iso & 512 $\times$ 512 & 15000 \\
D8  & 2.1, 3.5, 4.9, 5.9 & 100  & $120^\circ$, Iso & 512 $\times$ 512 & 15000 \\
D9  & 2.1, 3.3, 3.5, 4.9 & 120  & $30^\circ, 120^\circ$ & 512 $\times$ 512 & 15000 \\
D10 & 2.1, 3.3, 3.5, 4.9, 5.9 & 130  & $30^\circ, 120^\circ$, Iso & 512 $\times$ 512 & 15000 \\
\midrule
P1  & 2.1, 3.3, 3.5, 4.9, 5.9 & 150  & $30^\circ, 120^\circ$, Iso & 256 $\times$ 256 & 15000 \\
F1  & 2.6, 7.1 & 120  & $30^\circ, 120^\circ$, Iso & 512 $\times$ 512 & 15000 \\
A1  & 2.1, 3.3, 3.5, 4.9, 5.9 & 120  & $60^\circ$ & 512 $\times$ 512 & 15000 \\
\botrule
\end{tabular}}
\end{table}

\begin{table}[t]
\vspace{-4pt}
\centering
\caption{Network parameters of FARM with different sizes.}
\label{tab:fmure_sizes}
\tiny
{\setlength{\tabcolsep}{1pt}
\renewcommand{\arraystretch}{1.5}
\begin{tabular}{>{\centering\arraybackslash}p{0.12\columnwidth}*{6}{>{\centering\arraybackslash}p{0.11\columnwidth}}>{\centering\arraybackslash}p{0.12\columnwidth}@{}}
\toprule
\textbf{Model} & \textbf{Enc.\ depth} & \textbf{Enc.\ width} & \textbf{Enc.\ heads} & \textbf{Dec.\ depth} & \textbf{Dec.\ width} & \textbf{Dec.\ heads} & \textbf{Parameters} \\
\midrule
FARM-small  & 6  & 256 & 8 & 4 & 256 & 8 & 14.46M \\
FARM-base   & 8  & 512 & 8 & 6 & 512 & 8 & 63.93M \\
FARM-large  & 10 & 768 & 8 & 8 & 768 & 8 & 172.04M \\
\botrule
\end{tabular}}
\end{table}

\subsection{Benchmarks}

To comprehensively evaluate FARM across condition-free and condition-based construction tasks under three input cases, namely condition-free, condition-only, and condition-and-sample, we compare it against six representative benchmarks:

\begin{itemize}
\item \textbf{Kriging \cite{SaFu_TCCN_2017}:} A classical geo-statistical interpolation method that serves as a traditional baseline for condition-free ARM construction.

\item \textbf{AE \cite{TeRo_TWC_2022}:} A deep learning baseline for condition-free construction that employs an AE architecture to reconstruct spatial radio propagation patterns using only sparse observations.

\item \textbf{RadioDiff \cite{WTCY_TCCN_2025}:} A diffusion-driven benchmark representing the SOTA in generative modeling for condition-based construction under the condition-only input.

\item \textbf{RadioUNetC \cite{LYKC_TWC_2021}:} A representative image-to-image benchmark under the condition-only input, leveraging radio environmental priors to generate radio maps.

\item \textbf{RadioUNetS \cite{LYKC_TWC_2021}:} A variant of RadioUNetC that incorporates sparse observations as additional inputs, serving as a benchmark under the condition-and-sample input.

\item \textbf{RME-GAN \cite{ZWD_IoTJ_2023}:} A GAN-driven benchmark under the condition-and-sample input, which reconstructs radio maps by combining priors with estimates inferred from sparse observations.
\end{itemize}

Since these benchmarks are originally designed for terrestrial radio maps, we adapt them to aerial scenarios by splitting the ARM into height-wise slices for training. During inference, the predicted slices are stacked to reconstruct the ARM for metrics computation. Consistent with the unified nature of FARM, these benchmarks represent both traditional interpolation and advanced deep generative paradigms, ensuring a robust comparison across all operating modes.

\subsection{Performance Metrics}

To comprehensively evaluate the quality of the ARM construction, we adopt two widely used metrics for overall error measurement: normalized mean squared error (NMSE) and root mean square error (RMSE). Furthermore, since accurately generating structural details across the aerial space is crucial for network planning, we introduce peak signal-to-noise ratio (PSNR) and structural similarity index measure (SSIM) to evaluate the structural integrity and fidelity of the constructed ARMs. These metrics help assess whether the high-resolution granularity provided by ARM-Omni is preserved, allowing for a rigorous assessment of the model's ability to construct complex volumetric propagation patterns.

\subsubsection{NMSE and RMSE}
NMSE and RMSE measure the voxel-wise numerical precision of the ARM construction. These metrics are calculated based on the mean squared error (MSE) between the ground-truth ARM $\mathbf{R}$ and the estimated ARM $\hat{\mathbf{R}}$, which is defined as:
\begin{equation}
    \text{MSE} = \frac{1}{LWH} \sum_{l=1}^{L}\sum_{w=1}^{W}\sum_{h=1}^{H} \big(\mathbf{R}[l,w,h] - \hat{\mathbf{R}}[l,w,h]\big)^2.
\end{equation}
The RMSE is the square root of the MSE (i.e., $\text{RMSE} = \sqrt{\text{MSE}}$), while the NMSE scales the overall squared error by the total energy of the ground truth:
\begin{equation}
    \text{NMSE} = \frac{\sum_{l,w,h} \big(\mathbf{R}[l,w,h] - \hat{\mathbf{R}}[l,w,h]\big)^2}{\sum_{l,w,h} \big(\mathbf{R}[l,w,h]\big)^2}.
\end{equation}

\subsubsection{PSNR}
PSNR assesses the fidelity of ARM construction with an emphasis on edge preservation and local detail recovery. It evaluates the ratio between the maximum possible power of the signal and the power of the corrupting noise, defined as:
\begin{equation}
    \text{PSNR} = 10 \log_{10} \left( \frac{r^2}{\text{MSE}} \right),
\end{equation}
where $r$ represents the maximum variation range of the signal strength in the dataset. A higher PSNR indicates superior construction quality and minimal local distortions.

\subsubsection{SSIM}
SSIM evaluates the preservation of structural information, making it highly suitable for assessing texture and high-frequency variations in ARMs. Given two corresponding local 3D volumetric windows $\mathbf{x}$ and $\mathbf{y}$ from the ground-truth ARM and the estimated ARM, respectively, their structural similarity is computed as:
\begin{equation}
    \text{SSIM}(\mathbf{x}, \mathbf{y}) = \frac{(2\mu_x \mu_y + C_1)(2\sigma_{xy} + C_2)}{(\mu_x^2 + \mu_y^2 + C_1)(\sigma_x^2 + \sigma_y^2 + C_2)},
\end{equation}
where $\mu_x$ and $\mu_y$ denote the local means, $\sigma_x^2$ and $\sigma_y^2$ represent the local variances, and $\sigma_{xy}$ is the local covariance. The terms $C_1 = (K_1 r)^2$ and $C_2 = (K_2 r)^2$ with $K_1=0.01$, $K_2=0.03$ are constants included to maintain numerical stability. The SSIM is averaged across the entire volume to provide a global structural fidelity score.

\begin{figure}[!t]
\centering
\includegraphics[width=0.88\columnwidth]{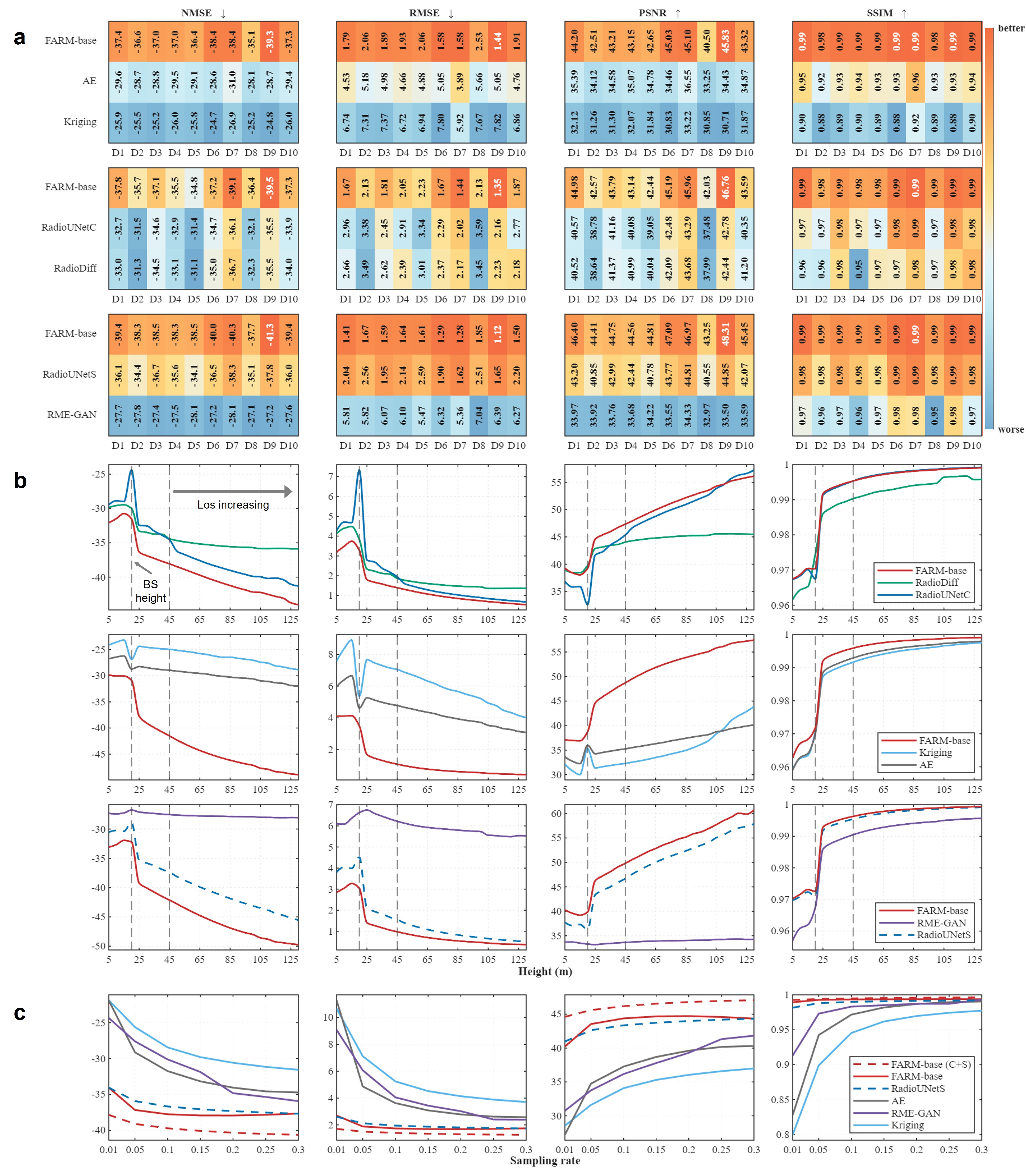}
\caption{\textbf{Overview of performance results of FARM and benchmarks.}
\textbf{a}. Performance comparison on D1--D10 under the condition-free input (top), condition-only input (middle), and condition-and-sample input (bottom).
\textbf{b}. Height-wise performance comparison under the condition-only input (top), condition-free input (middle), and condition-and-sample input (bottom).
\textbf{c}. Performance comparison of the condition-free and condition-and-sample input cases.
}
\label{fig:comparsion_results}
\end{figure}

\subsection{Performance Evaluation}

\subsubsection{Multi-dataset Unified Learning}
\label{Multi_dataset_unified_earning}

The in-domain evaluation results, summarized in Fig.~\ref{fig:comparsion_results}a, demonstrate that FARM achieves SOTA performance across all ten datasets. In the context of condition-free construction, AE emerges as the strongest baseline. Compared to AE, FARM reduces the NMSE by an average of $8.13$ dB and the RMSE by $2.99$ dB while simultaneously improving PSNR and SSIM by $8.81$ dB and $5\%$, respectively. When compared to Kriging, these advantages are even more pronounced, with average NMSE and RMSE reductions of $11.66$ dB and $5.24$ dB. These significant gains highlight the superior ability of FARM to enhance overall construction accuracy and preserve structural fidelity within the aerial radio environment. 

For the condition-only input of the condition-based construction task, RadioUNetC and RadioDiff provide comparable baseline performance. FARM outperforms RadioUNetC by decreasing the average NMSE and RMSE by $3.52$ dB and $0.95$ dB, respectively, alongside gains of $3.44$ dB in PSNR and $1\%$ in SSIM. Similar improvements are observed over RadioDiff, where FARM reduces the average NMSE by $3.40$ dB and RMSE by $0.81$ dB. Notably, although RadioDiff employs an advanced generative formulation, it does not significantly outperform RadioUNetC in this aerial context. This performance plateau is attributed to the fact that both benchmarks operate in a slice-wise manner, which introduces data distribution heterogeneity across altitude slices. Specifically, RadioUNetC benefits from the deterministic mapping provided by the FSPL channel of the input, allowing it to capture propagation patterns more effectively in high-altitude slices, which account for the majority of the data. By contrast, while RadioDiff's stochastic denoising process enhances construction performance in low-altitude slices, it also increases the sensitivity to distribution heterogeneity, thereby reducing the accuracy for high-altitude slices. Consequently, RadioUNetC achieves higher average performance than RadioDiff. Under the condition-and-sample input, RadioUNetS outperforms RME-GAN across all metrics because the input samples directly provide the additional supervised signal for construction. By contrast, RME-GAN first estimates a radio map from the samples using a traditional method and then adopts the estimated map as supervision, which inevitably introduces estimation noise and degrades the construction accuracy. Nevertheless, despite the strong performance of RadioUNet variants in individual operating modes, they require separate retraining for different inputs. In contrast, FARM leverages its foundation model architecture to model aerial radio propagation features, enabling consistent improvements over baselines across all operating modes.

\subsubsection{Altitude Analysis}

As illustrated in Fig.~\ref{fig:comparsion_results}b, all methods exhibit the highest construction error at lower altitudes, where dense blockages and severe multipath fading make the propagation environment particularly challenging. Their performance improves above $45$ m as dominant LoS links begin to emerge. In condition-based tasks, RadioUNetC and RadioUNetS show strong performance in high-altitude slices and achieve an NMSE below $-35$ dB under both condition-only and condition-and-sample inputs, respectively. These results highlight the advantage of deterministic mapping under relatively regular LoS-dominated conditions. RadioDiff also benefits from the more regular propagation conditions at higher altitudes, but its performance tends to saturate because of the stochastic nature of its denoising process. In contrast, RME-GAN relies heavily on estimated maps that remain similar across altitudes, so its performance does not show a clear improvement as altitude increases. These results further support the analysis in \ref{Multi_dataset_unified_earning}. In condition-free tasks, AE achieves better construction performance compared with Kriging over most altitude ranges, particularly in complex near-ground regions. However, as altitude increases, Kriging improves more markedly and even surpasses AE in PSNR above $105$ m, which indicates that traditional interpolation is more suited to capturing the smooth large-scale trends in high-altitude LoS-dominated regions. By contrast, FARM remains more robust in both condition-based and condition-free settings. This advantage arises from its ability to capture 3D dependencies more effectively than slice-wise approaches, resulting in the most balanced performance across the full height range.

A notable exception occurs at the BS altitude of $20$ m, where the radio map contains the largest proportion of no-signal regions as shown in Fig. \ref{fig:dataset}b and therefore exhibits a spatial distribution distinct from that of other height slices. This altitude-specific distribution shift affects condition-based and condition-free tasks differently. Under the condition-only input setting, the conditioning information cannot fully capture this shift, thereby creating a mismatch between the conditions and the target that leads to a local error spike. This effect is particularly evident in RadioUNet, whose image-to-image formulation appears more sensitive to such inconsistencies. In contrast, condition-free methods estimate the radio map directly from sparse samples, which already convey the distribution prior of this height slice. Therefore, large no-signal regions are reconstructed more faithfully, leading to a local error trough at this altitude. The results under the condition-and-sample input further support this interpretation, as the error spike is reduced compared with the condition-only mode after sparse samples are introduced. Although FARM is also affected by the altitude-dependent shift, its error variations are much milder in both construction settings due to its more effective modeling of altitude-dependent distribution characteristics.

\subsubsection{Robustness to Sampling Rate}

We next evaluate whether FARM can remain reliable when only a very limited number of RSS samples are available. For the condition-free task and the condition-and-sample input case, the sampling rate is varied only during testing, without retraining or adapting any model parameters. As shown in Fig.~\ref{fig:comparsion_results}c, all methods improve as more samples are provided, but their sensitivity to sampling density is markedly different. The performance of AE and Kriging drops rapidly under extremely sparse observations, indicating that the sparse samples alone are insufficient for them to recover the global propagation structure. In particular, when the sampling rate decreases to $1\%$, their NMSE values are only around $-22$ dB, whereas FARM still achieves approximately $-34$ dB in the condition-free task.

This robustness mainly stems from the fact that FARM does not rely on local sample interpolation alone. Instead, the radio encoder extracts high-level spatial representations from the visible RSS patches, while the map decoder completes the missing regions according to the propagation priors learned during large-scale pretraining. Therefore, even when the observed samples are too sparse to describe the full field explicitly, they can still serve as anchors for selecting a physically plausible radio-map distribution. A similar trend is observed under the condition-and-sample input case, where FARM achieves an NMSE of approximately $-37$ dB at the $1\%$ sampling rate and consistently outperforms RadioUNetS and RME-GAN. This also indicates that sparse samples and environmental conditions are complementary in FARM. Overall, these results show that FARM is less sensitive to measurement scarcity and can support high-fidelity ARM construction in practical deployments where dense radio measurements are costly or unavailable.

\begin{figure}[!t]
\centering
\includegraphics[width=0.88\columnwidth]{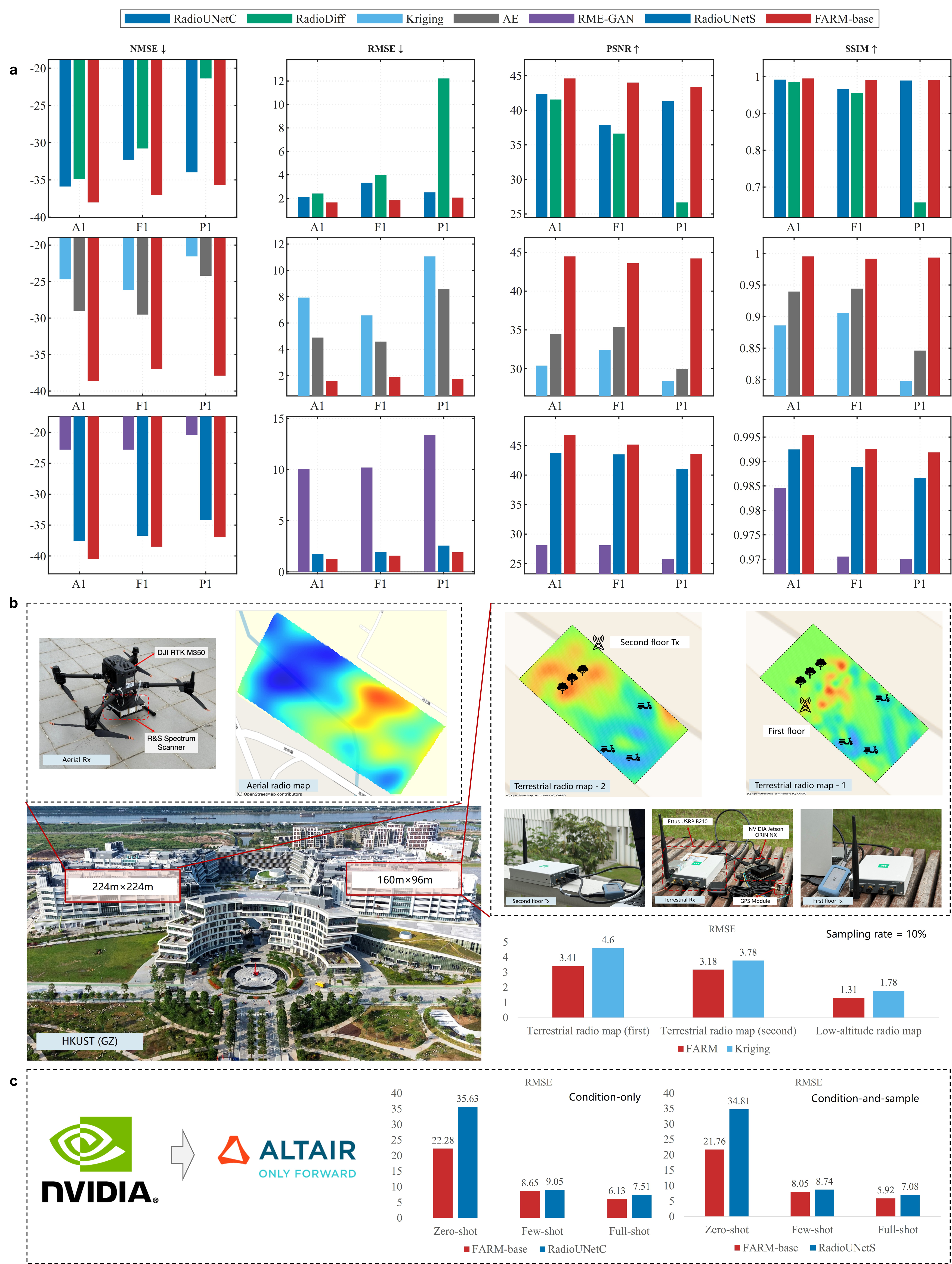}
\caption{\textbf{Overview of generalization performance for FARM and benchmark methods.}
\textbf{a}. Zero-shot performance comparison on A1, F1, and P1 under the condition-free input (top), condition-only input (middle), and condition-and-sample input (bottom).
\textbf{b}. Transferability performance of FARM on the real-world dataset (HKUST Guangzhou Campus).
\textbf{c}. Transferability performance of FARM on the ALTAIR simulation dataset (RadioMapSeer).
}
\label{fig:generation_results}
\end{figure}

\subsubsection{Zero-shot Generalization}

Fig.~\ref{fig:generation_results}a presents the zero-shot comparison on the unseen subsets P1, F1, and A1. All methods are evaluated directly on the target set without adaptation, where FARM uses the trained FARM-base model and each benchmark reports the average performance of ten source-domain models. These subsets test generalization under three types of distribution shifts. Specifically, F1 introduces unseen carrier frequencies with different propagation characteristics, while A1 uses a new antenna pattern that changes the spatial energy distribution of ARMs. P1 extends the height range while reducing the horizontal coverage to one-quarter of the original area, thereby testing spatial transferability across unseen altitude ranges and LoS transitions.

Across all three distribution shifts, FARM exhibits superior generalization capability in all task settings. In the condition-free task, AE generalizes better than Kriging because it can exploit learned structural priors, whereas Kriging is limited to local interpolation and struggles to recover large-scale field variations in unseen environments. Nevertheless, FARM substantially outperforms AE by leveraging the unified aerial representation extracted by the radio encoder, achieving PSNR gains of $9.99$ dB, $8.23$ dB, and $14.21$ dB on A1, F1, and P1, respectively. Under the condition-only input, RadioUNetC generalizes better than RadioDiff, suggesting that its deterministic mapping is more stable under distribution shifts than the multi-step stochastic denoising process of RadioDiff. This advantage is especially clear on P1, where the spatial coverage and LoS transitions differ from the source domains. However, RadioUNetC still suffers under the unseen frequency shift F1, indicating that direct image-to-image mapping remains tied to source-domain propagation patterns and is less flexible when the underlying radio characteristics change. FARM addresses this limitation through its flow-based map decoder. The decoder replaces direct mapping with a deterministic denoising process, allowing the model to adapt to spatial structure changes with only one or two additional denoising steps while reducing its dependence on input conditions. Under the condition-and-sample input, RME-GAN generalizes poorly due to its dependence on estimated results, whereas RadioUNetS uses sparse samples to correct deterministic mapping bias and improve generalization. However, its performance remains consistently below FARM, suggesting that directly entangling sparse-sample understanding with condition-based mapping is not optimal under distribution shifts. By decoupling representation extraction from condition-guided construction, FARM achieves stronger construction performance and the best generalization across A1, F1, and P1. Additionally, the antenna-pattern shift A1 is relatively less challenging for most methods, as the narrower beamwidth produces larger no-signal regions and thus a simpler spatial structure than the isotropic antenna case. Overall, these results demonstrate that FARM achieves more comprehensive and robust generalization, providing a reliable foundation for high-fidelity ARM construction in low-altitude economy applications.

\subsubsection{Transferability Evaluation}

To further assess the transferability of FARM beyond ARM-Omni, we evaluate it on two external datasets: a real-world dataset collected on the HKUST Guangzhou campus and the representative terrestrial radio map dataset RadioMapSeer simulated using ALTAIR. Specifically, Fig.~\ref{fig:generation_results}b shows the collection regions and measurement procedure of the real-world dataset. We collect a low-altitude radio map over a $224\,\mathrm{m} \times 224\,\mathrm{m}$ area at $100\,\mathrm{m}$ using a DJI RTK M350 equipped with a Rohde \& Schwarz spectrum scanner, as well as two terrestrial radio maps over a $160\,\mathrm{m} \times 96\,\mathrm{m}$ area using an NVIDIA Jetson Orin NX and an Ettus USRP B210, where the transmitter is placed on the first and second floors, respectively. Since accurate environmental maps and BS configurations are unavailable, and the dataset size is insufficient for fine-tuning or retraining, FARM is evaluated in the condition-free mode under zero-shot transfer. Because FARM uses a patch size of two along the height dimension, each single-layer radio map is duplicated and stacked into a two-layer input during testing. At a sampling rate of $10\%$, FARM achieves RMSE values of $3.41$ dB, $3.18$ dB, and $1.31$ dB on the first-floor terrestrial map, second-floor terrestrial map, and low-altitude map, respectively. The lower RMSE on the low-altitude map indicates that FARM performs better in a LoS-dominated scenario. Compared with Kriging, FARM reduces RMSE by $1.19$ dB, $0.60$ dB, and $0.47$ dB on the three maps, respectively. The largest improvement appears in the more NLoS-dominated first-floor scenario, where interpolation is easily affected by blockage-induced nonstationarity. This result suggests that FARM can exploit learned representations beyond local spatial smoothness, demonstrating its practical transferability to real-world radio map construction.

We further evaluate FARM on RadioMapSeer to examine the cross-simulator transfer from Sionna-based ARM-Omni to ALTAIR-generated radio maps, as shown in Fig.~\ref{fig:generation_results}c. FARM is tested under both condition-only and condition-and-sample inputs.  In the zero-shot setting, FARM outperforms RadioUNet by reducing the RMSE from $35.63$ dB to $22.28$ dB under the condition-only input and from $34.81$ dB to $21.76$ dB under condition-and-sample input. The remaining high errors, however, indicate a substantial simulator gap between the two propagation engines. Few-shot adaptation with only $20\%$ of the RadioMapSeer data sharply reduces the RMSE to $8.65$ dB and $8.05$ dB under the two input settings, respectively. With full target-domain training, FARM further improves to $6.13$ dB and $5.92$ dB, consistently outperforming RadioUNet. These results suggest that the propagation priors learned from Sionna-based simulations are transferable to ALTAIR scenarios, but effective transfer still benefits from limited target-domain calibration. The observed gap is likely caused by differences in multipath modeling, reflection, diffraction, scattering, and material configurations across simulators, highlighting the need to incorporate more diverse simulated and real-world radio maps to improve FARM's transferability.

\begin{table}[t]
\centering
\caption{Average performance of FARM under different model sizes and input cases. The \colorbox{red!15}{Light red} shading indicates the best result in each metric group.}
\label{tab:model_scaling_results}
\tiny
{\setlength{\tabcolsep}{1pt}
\renewcommand{\arraystretch}{1.5}
\begin{tabular}{>{\centering\arraybackslash}p{0.20\columnwidth}>{\centering\arraybackslash}p{0.15\columnwidth}*{4}{>{\centering\arraybackslash}p{0.1375\columnwidth}}@{}}
\toprule
\textbf{Input} & \textbf{Model} & \metricdown{NMSE} & \metricdown{RMSE} & \metricup{PSNR} & \metricup{SSIM} \\
\midrule
\multirow{3}{=}{\centering\textbf{Condition-free}} & FARM-small & -36.25 & 2.13 & 42.23 & 0.9914 \\
                                                   & FARM-base  & -37.17 & 1.88 & 43.55 & 0.9928 \\
                                                   & FARM-large & \cellcolor{red!15}{\textbf{-37.55}} & \cellcolor{red!15}{\textbf{1.80}} & \cellcolor{red!15}{\textbf{44.06}} & \cellcolor{red!15}{\textbf{0.9935}} \\
\cmidrule{1-6}
\multirow{3}{=}{\centering\textbf{Condition-only}} & FARM-small & -35.34 & 2.15 & 42.69 & 0.9908 \\
                                                   & FARM-base  & -36.85 & 1.84 & 44.05 & 0.9920 \\
                                                   & FARM-large & \cellcolor{red!15}{\textbf{-37.17}} & \cellcolor{red!15}{\textbf{1.81}} & \cellcolor{red!15}{\textbf{44.10}} & \cellcolor{red!15}{\textbf{0.9923}} \\
\cmidrule{1-6}
\multirow{3}{=}{\centering\textbf{Condition-and-sample}} & FARM-small & -38.32 & 1.63 & 44.56 & 0.9917 \\
                                                         & FARM-base  & -39.17 & 1.49 & 45.70 & 0.9928 \\
                                                         & FARM-large & \cellcolor{red!15}{\textbf{-39.34}} & \cellcolor{red!15}{\textbf{1.41}} & \cellcolor{red!15}{\textbf{46.01}} & \cellcolor{red!15}{\textbf{0.9931}} \\
\botrule
\end{tabular}}
\end{table}

\begin{table}[t]
\centering
\caption{Comparison of model size and inference efficiency.}
\label{tab:complexity_results}
\tiny
{\setlength{\tabcolsep}{1pt}
\renewcommand{\arraystretch}{1.5}
\begin{tabular}{>{\centering\arraybackslash}p{0.20\columnwidth}*{4}{>{\centering\arraybackslash}p{0.175\columnwidth}}@{}}
\toprule
\textbf{Method} & \textbf{Params (M)} & \textbf{Heights / pass} & \textbf{Time / pass (s)} & \textbf{Time / height (s)} \\
\midrule
Kriging          & N/A    & 1  & 0.0043 & 0.0043 \\
AE               & 37.58  & 1  & 0.0067 & 0.0067 \\
RadioUNet        & 13.27  & 1  & 0.0041 & 0.0041 \\
RadioDiff        & 297.74 & 1  & 0.5553 & 0.5553 \\
RME-GAN          & 16.04  & 1  & 0.0045 & 0.0045 \\
FARM-base        & 63.93  & 30 & 0.1096 & 0.0037 \\
\botrule
\end{tabular}}
\end{table}

\subsubsection{Scalability and Deployment}

We further evaluate the scalability and deployment efficiency of FARM. Table~\ref{tab:model_scaling_results} compares FARM-small, FARM-base, and FARM-large under the three input cases. A clear scaling trend can be observed: increasing model capacity consistently improves metrics, indicating that larger FARM variants can learn richer radio representations rather than benefiting only under a specific input. Meanwhile, the improvement becomes less pronounced when scaling from FARM-base to FARM-large. Specifically, the NMSE gains from FARM-base to FARM-large are only $0.38$ dB, $0.32$ dB, and $0.17$ dB under the condition-free, condition-only, and condition-and-sample inputs, respectively. These diminishing returns suggest that FARM-base already captures most of the relevant propagation structure, while FARM-large mainly provides additional refinement.

We then evaluate deployment efficiency using FARM-base, which provides a representative balance between model capacity and reconstruction accuracy. Table~\ref{tab:complexity_results} compares the deployment cost of different methods. FARM-base has $63.93$M parameters and requires $0.1096$ s for one forward pass, which is higher than most slice-wise baselines. However, unlike these methods that construct only one height slice per pass, FARM predicts all $30$ height slices simultaneously. As a result, its normalized inference time is only $0.0037$ s per height, the lowest among all compared methods. Overall, these results indicate that FARM not only scales effectively in reconstruction accuracy but also remains efficient for high-dimensional ARM construction in practical deployment.

\begin{table}[t]
\centering
\caption{Average performance gain brought by map decoder fine-tuning. $\Delta$ denotes the performance improvement.}
\label{tab:finetuning_gain}
\tiny
{\setlength{\tabcolsep}{1pt}
\renewcommand{\arraystretch}{1.5}
\begin{tabular}{>{\centering\arraybackslash}p{0.20\columnwidth}>{\centering\arraybackslash}p{0.15\columnwidth}*{4}{>{\centering\arraybackslash}p{0.1375\columnwidth}}@{}}
\toprule
\textbf{Input} & \textbf{Model} & $\Delta$\textbf{NMSE} & $\Delta$\textbf{RMSE} & $\Delta$\textbf{PSNR} & \textbf{$\Delta$SSIM} \\
\midrule
\multirow{3}{=}{\centering\textbf{Condition-free}} & FARM-small & 0.47 & 0.12 & 0.50 & 0.00022 \\
                                                   & FARM-base  & 0.26 & 0.07 & 0.46 & -0.00001 \\
                                                   & FARM-large & 0.28 & 0.07 & 0.51 & 0.00016 \\
\cmidrule{1-6}
\multirow{3}{=}{\centering\textbf{Condition-only}} & FARM-small & 7.09 & 2.97 & 7.46 & 0.01860 \\
                                                   & FARM-base  & 15.13 & 8.24 & 13.08 & 0.02082 \\
                                                   & FARM-large & 8.34 & 2.92 & 7.90 & 0.02095 \\
\cmidrule{1-6}
\multirow{3}{=}{\centering\textbf{Condition-and-sample}} & FARM-small & 0.31 & 0.11 & 0.45 & 0.00012 \\
                                                         & FARM-base  & 0.18 & 0.05 & 0.41 & 0.00007 \\
                                                         & FARM-large & 0.16 & 0.05 & 0.43 & 0.00001 \\
\botrule
\end{tabular}}
\end{table}

\begin{table}[t]
\centering
\caption{Ablation of the BS location channel in the condition inputs. The \colorbox{red!15}{Light red} shading indicates the best result in each metric group.}
\label{tab:bs_location_channel}
\tiny
{\setlength{\tabcolsep}{1pt}
\renewcommand{\arraystretch}{1.5}
\begin{tabular}{>{\centering\arraybackslash}p{0.27\columnwidth}>{\centering\arraybackslash}p{0.15\columnwidth}*{4}{>{\centering\arraybackslash}p{0.088\columnwidth}}>{\centering\arraybackslash}p{0.14\columnwidth}@{}}
\toprule
\textbf{Input} & \textbf{BS Location} & \metricdown{NMSE} & \metricdown{RMSE} & \metricup{PSNR} & \metricup{SSIM} & \metricdown{Loc. err. (m)} \\
\midrule
\multirow{2}{=}{\centering\textbf{Condition-only}} & w/o BS loc. & -36.22 & 2.14 & 43.59 & 0.98 & 19.74 \\
                                                     & w/ BS loc.  & \cellcolor{red!15}{\textbf{-37.09}} & \cellcolor{red!15}{\textbf{1.84}} & \cellcolor{red!15}{\textbf{44.05}} & \cellcolor{red!15}{\textbf{0.99}} & \cellcolor{red!15}{\textbf{19.11}} \\
\cmidrule{1-7}
\multirow{2}{=}{\centering\textbf{Condition-and-sample}} & w/o BS loc. & -38.36 & 1.67 & 44.42 & 0.99 & 19.70 \\
                                                         & w/ BS loc.  & \cellcolor{red!15}{\textbf{-39.22}} & \cellcolor{red!15}{\textbf{1.50}} & \cellcolor{red!15}{\textbf{45.61}} & \cellcolor{red!15}{\textbf{0.99}} & \cellcolor{red!15}{\textbf{19.27}} \\
\botrule
\end{tabular}}
\end{table}

\subsubsection{Ablation Experiments}

We conduct two ablation studies to clarify which components are responsible for FARM's unified construction capability. The first ablation examines the role of generative fine-tuning for the map decoder, as reported in Table~\ref{tab:finetuning_gain}. Its effect is strongly input-dependent. In the sample-involved settings, the gains are consistently small. This indicates that the pretrained radio encoder already extracts effective radio representations from sparse RSS observations, so the decoder mainly performs representation-to-map refinement. By contrast, generative fine-tuning becomes much more important under the condition-only input, where no RSS samples are available and the ARM must be generated from the radio environmental conditions alone. In this case, fine-tuning substantially improves all model sizes. These results show that pretraining provides FARM with strong radio representation ability, whereas decoder generative fine-tuning is essential for conditional construction ability.

The second ablation evaluates the necessity of the explicit BS location channel, since the FSPL already contains transmitter-distance information. For this purpose, FARM-base is retrained without the BS location channel while keeping the other condition inputs unchanged. To quantify its impact more comprehensively, we introduce a new metric, namely location error, defined as the Euclidean distance between the true BS location and the location inferred from the constructed ARM by identifying the voxel with the maximum reconstructed RSS. As shown in Table~\ref{tab:bs_location_channel}, removing the explicit BS location channel leads to consistent degradation. Under the condition-only input, NMSE degrades from $-37.09$ dB to $-36.22$ dB, RMSE increases from $1.84$ dB to $2.14$ dB, and location error increases from $19.11$ m to $19.74$ m. Under the condition-and-sample input, the same removal changes NMSE from $-39.22$ dB to $-38.36$ dB and increases location error from $19.27$ m to $19.70$ m. These results indicate that the FSPL alone cannot fully replace the explicit BS location channel. Although FSPL provides a continuous geometric prior, the BS location offers a direct spatial anchor for identifying the transmitter-centered propagation structure. Therefore, the explicit BS location channel is necessary for improving both reconstruction accuracy and BS localization consistency, and is retained as a necessary condition input in FARM.

\section{Conclusion}

This paper developed and validated FARM, the first unified foundation model for ARM construction in low-altitude networks. By coupling a MAE with a diffusion-based decoder in a dual-stage architecture, FARM achieved both condition-free and condition-based ARM construction within a single framework, enabling flexible use of sparse RSS measurements and radio environmental priors. To support FARM training, we constructed ARM-Omni, a large-scale, high-granularity 3D ARM dataset with multi-band and multi-antenna configurations, which captured fine-grained spatial variations across both horizontal and vertical dimensions. Extensive experiments demonstrated that FARM consistently outperformed SOTA benchmarks and maintained strong generalization across varying coverage dimensions, carrier frequencies, and antenna patterns. Transfer evaluations on real-world field-test data further validated the practical deployment capability of FARM. Ultimately, this work provides a fully verified technological foundation to support the low-altitude economy.

\bibliography{cite}

\end{document}